\documentclass[useAMS,usenatbib,ps2pdf]{mn2e}

\pdfoutput=1
\usepackage{graphicx}
\usepackage{deluxetable}
\usepackage{amssymb}
\usepackage{color}

\title[]{Globular Cluster Systems in Nearby Dwarf Galaxies: III. Formation
  Efficiencies of Old Globular Clusters\thanks{Partly based on archival
    data of the NASA/ESA {\it Hubble Space Telescope}, which is operated by
    AURA, Inc., under NASA contract NAS 5--26555.}} 

\author[I.\,Y.\,Georgiev et al.]{Iskren
  Y. Georgiev$^{1}$\thanks{E-mail: iskren@astro.uni-bonn.de}, Thomas
  H. Puzia$^{2}$, Paul Goudfrooij$^{3}$ and Michael Hilker$^{4}$\\ 
$^{1}$Argelander Institut f\"{u}r Astronomie der Universit\"{a}t Bonn,
Auf dem H\"{u}gel 71, D-53121 Bonn, Germany\\ 
$^{2}$Herzberg Institute for Astrophysics, National Research Council, 5071 West
Saanich Road, Victoria, BC V9E 2E7, Canada\\ 
$^{3}$Space Telescope Science Institute, 3700 San Martin Drive,
Baltimore, MD 21218, USA\\
$^{4}$European Southern Observatory, Karl-Schwarzschild-Str.\,2, 85748
Garching bei M\"{u}nchen, Germany
}
\begin{document}

\date{April 2010}

\pagerange{\pageref{firstpage}--\pageref{lastpage}} \pubyear{2009}

\maketitle

\label{firstpage}

\begin{abstract}
We investigate the origin of the shape of the globular cluster (GC) system 
scaling parameters as a function of galaxy mass, i.e. specific frequency
($S_N$), specific luminosity ($S_L$), specific mass ($S_M$), and specific 
number ($\hat{T}$) of GCs.  In the low-mass galaxy regime ($M_V\!\ga\!-16$ 
mag) our analysis is based on {\it HST/ACS} observations of GC populations 
of faint, mainly late-type dwarf galaxies in low-density environments. In 
order to sample the entire range in galaxy mass ($M_V\!=\!-11$ to $-23$\,mag 
$=10^6\!-\!10^{11}L_\odot$), environment, and morphology we augment our sample 
with data of spiral and elliptical galaxies from the literature, in which 
old GCs are reliably detected. This large dataset confirms (irrespective 
of galaxy type) the increase of the specific frequencies of GCs above and 
below a galaxy magnitude of $M_V\simeq-20$\,mag. Over the full mass range, 
the $S_L-$value of early-type galaxies is, on average, twice that of late-types. 
To investigate the observed trends we derive theoretical predictions of 
GC system scaling parameters as a function of host galaxy mass based on 
the models of \cite{Dekel&Birnboim06} in which star-formation processes 
are regulated by stellar and supernova feedback below a stellar mass of 
$3\!\times\!10^{10}{\cal M}_\odot$, and by virial shocks above it. We find 
that the analytical model describes remarkably well the shape of the GC system 
scaling parameter distributions with a  universal {\it specific GC formation 
efficiency}, $\eta$, which relates the total mass in GCs to the total galaxy 
halo mass. Early-type and late-type galaxies show a similar mean value of 
$\eta\!=\!5.5\times10^{-5}$, with an increasing scatter towards lower galaxy 
masses. This 
can be due to the enhanced stochastic nature of the star and star-cluster 
formation processes for such systems. Some massive galaxies have excess $\eta$ 
values compared to what is expected from the mean model prediction for galaxies 
more luminous than $M_V\simeq-20$\,mag ($L_V\gtrsim10^{10}L_\odot$). This may 
be attributed to a very efficient early GC formation, less efficient production 
of field stars or accretion of predominantly low-mass/luminosity high$-\eta$ 
galaxies, or a mixture of all these effects.
\end{abstract}


%%%%%%%%%%%%%%%%%%%%%%%%%%%%%%%%%%%%%%%
\section{Introduction}\label{intro}

\vspace{4mm}One of the very first stellar systems to form in the early 
Universe are old globular clusters (GCs), which are observed in galaxies 
of all morphological types. Globular clusters are massive agglomerations 
of gravitationally bound stars, the majority of which formed almost 
simultaneously out of gas with similar chemical composition. As fossil
records of the first star formation episodes of their host galaxies, the
distribution of their integrated properties (age, mass, metallicity,
structural parameters) as well as the general properties of all GCs in a
galaxy (total numbers, spatial and dynamical distributions) hold important
clues to the initial physical conditions at which they have formed and
evolved. For that reason, the global properties of globular cluster
systems (GCSs) have long been recognized as promising tools to study the
major galaxy star formation episodes and to serve as observational
constraints to differentiate between various models of galaxy formation 
\protect{\cite[][and references therein]{Kissler-Patig00,vdBergh00, 
Harris91,Harris03,Brodie06}}.

One of the most commonly used parameters to describe GCSs is the
\emph{specific frequency} ($S_N$), i.e. the number of GCs per unit galaxy
luminosity. In essence, $S_N$ measures the formation efficiency of GCs 
relative to field stars. GCs and field stars are linked through the 
dissolution of star clusters due to various mechanisms (e.g. tidal shocks, 
cluster relaxation), which shapes the initial power-law to the observed, 
present-day Gaussian globular cluster luminosity function \cite[GCLF, 
see e.g.][and references therein]{Fall&Zhang01, Goudfrooij04, Goudfrooij07, 
Gieles06, McLaughlin&Fall08, Kruijssen&Zwart09, Elmegreen10}. The $S_N$ 
parameter was initially introduced by \cite{Harris&vdBergh81} as a measure 
of the richness of GCSs in elliptical galaxies. Since then $S_N$ has been 
applied by numerous studies to galaxies of different morphological types
from early-type, quiescent ellipticals to actively star-forming, late-type
spirals, irregulars, and interacting/merger galaxies, covering the entire
range in galaxy mass (from giants to dwarfs) and environments (from
galaxies in dense clusters to such in loose groups and in the field).
Wide-field ground-based and deep Hubble Space Telescope (HST) studies show that
the $S_N$ value varies \emph{greatly} among galaxies, particularly
among the most luminous ellipticals and low-mass dwarf galaxies 
\cite[e.g.][]{Harris91,Harris01,Wehner08,Peng08}. Spiral galaxies tend to
show much less scatter in their $S_N$, with values in the range $0.5-2$
\citep{Ashman&Zepf98,Goudfrooij03,Chandar04,Rhode07}. The general trend
observed for early-type dwarf elliptical (dE) galaxies is that their average
$S_N$ value increases with decreasing galaxy luminosity from a few to a few tens 
\citep{Durrell96, Miller98,Miller&Lotz07,Peng08}. It
was shown recently that a similar behavior of $S_N$ holds for
late-type dwarfs \cite[dIrrs,][]{Seth04, Olsen04, Sharina05, Georgiev08,
Puzia&Sharina08}. Conversely, in the high-mass galaxy regime, a significant
increase of the $S_N$ values is observed for the 
most luminous galaxies \cite[gEs and cDs, e.g.][]{Rhode05, Peng08, Harris09}. 
As the stellar populations (colours and integrated light spectral indices) of
the latter galaxies are not significantly different from galaxies 1-2 mag
fainter, such an upturn in the $S_N$ distribution implies that the assembly
history of the most massive galaxies must have been different from those of 
less massive galaxies. 

The overall trend for $S_N$ is that with increasing galaxy luminosity from
$M_V\simeq-11$\,mag to $M_V\simeq-20$\,mag, $S_N$ decreases from the range 
$\sim0\!-\!100$ to $S_N\sim0.5\!-\!3$, respectively. For galaxies more 
luminous than $M_V\simeq-20$\,mag the $S_N$ increases again (to $\sim 10$).

To explain the increasing fraction of GCs over field stars in the most 
massive galaxies several studies suggested that the mass in GCs is proportional
to the total gas mass supply \cite[][]{West95,Blakeslee97, Blakeslee99}. 
In particular, \cite{McLaughlin99} investigated the $S_N$ behavior for the 
entire early-type galaxy mass range, from giants to dwarfs. He showed that 
the enhanced $S_N$ for the most massive galaxies considered in his sample 
(NGC\,1399, M\,87, M\,49) can be accounted for if the mass in GCs is 
normalized to the total baryonic mass of the host (stellar plus the hot 
X-ray emitting gas), implying a constant baryonic GC formation efficiency 
$\epsilon=0.26\%$. The parameter $\epsilon$ represents the efficiency of 
converting baryonic matter into GCs, $\epsilon\equiv{\cal M}_{\rm GCS}/{\cal M}_b$. 
However, as suggested by \cite{Blakeslee97} and \cite{Blakeslee99}, perhaps 
more fundamental is the ratio between the mass in GCs and the {\it total\/} 
galaxy halo mass ${\cal M}_h$, (i.e., $\eta_h\equiv{\cal M}_{\rm GCS}/{\cal M}_h$) 
for which they obtain $\eta_h\simeq1.71\times10^{-4}$. More recently, 
\cite{Kravtsov&Gnedin05} in a high-resolution cosmological simulation for 
galaxies with ${\cal M}_h  >10^9{\cal M_\odot}$ find ${\cal M}_{\rm GCS}=
3.2\times10^6{\cal M}_\odot ({\cal M}_h/10^{11}{\cal M}_\odot)^{0.13}$, which 
predicts $\epsilon_h=(2-5)\times10^{-5}$. Assuming a cosmological baryon 
fraction $f_b=0.17$ \citep{Hinshaw09} brings the observed and theoretically 
predicted values of the GC formation efficiency to a common value of $\sim10^{-5}$ 
(see Sect.\,\ref{Sect:GCefficiency}). Following up on the idea of the 
proportionality between the mass in GCs and the host halo mass using a 
statistical stellar-halo mass relation from $\Lambda$CDM simulations, recent 
studies with much larger galaxy samples find a good approximation to the 
data \citep{Peng08} and a similar result for the GC formation efficiency 
\cite[$\eta_h\!=\!7\times10^{-5}$,][]{Spitler09}. In the low-mass galaxy 
regime, \cite{Peng08} find evidence for an environmental bias: the majority 
of dwarf galaxies with relatively high $S_N$ at a given total luminosity 
lie within a projected radius of $\sim1$\,Mpc from M\,87, the central giant
elliptical in the Virgo galaxy cluster. The study of \cite{Miller&Lotz07} 
finds a consistent trend of increasing GC mass fraction with decreasing 
host galaxy mass for Virgo dwarf elliptical galaxies. Using the 
\cite{McLaughlin99} correction for ${\cal M}^{\rm init}_{\rm gas}/{\cal
M}^{\rm init}_{\rm star}$, \citeauthor{Miller&Lotz07} were able to match 
the trend seen in their observations.

Investigating the increasing $S_N$ with decreasing galaxy mass below
${\cal M}_\star\!<\!3\times10^{10}{\cal M}_\odot$, \cite{Forbes05} used 
the feedback models of \cite{Dekel&Woo03} which predict ${\cal M}/L\propto 
{\cal M}^{-2/3}$, and found qualitative agreement with the observations.
However, the normalization of this relation remained unconstrained, so that
actual GC formation efficiency values $\epsilon_h$ could not be determined. 
This quantitative derivation of GC system scaling relations as a function 
of galaxy mass is one of the goals of this work by including predictions 
from the latest \cite{Dekel&Birnboim06} models which include shock-heating 
regulated star formation for ${\cal M}_\star>3\times10^{10}{\cal M}_\odot$. 
In Section\,\ref{Sect:Data} we briefly introduce the galaxy sample used in 
this study and access contamination. The analysis of the fractions of GCs 
in late-type dwarf galaxies along with complementing data from the literature 
is presented in Section\,\ref{Sect:Analysis}. In Section\,\ref{Sect:Discussion}
we discuss the observed trends of the various GCS scaling relations, the 
specific GC formation efficiency $\eta$, and implications for galaxy formation 
scenarios. Combined with dynamical mass measurements of massive and dwarf galaxies 
from the literature we normalize the models (Sect.\,\ref{Sect:GCscalingNorm})
and find a good description of the observed behavior of the frequencies of GCs
as a function of galaxy luminosity with a common value of the GC formation
efficiency parameter. Using the derived analytical expressions describing 
the behavior of the specific GC system scaling parameters as a function of 
galaxy luminosity we discuss (in Sect.\,\ref{Sect:impl.4_h.g.f.}) the 
predictions of a simplistic satellite galaxy accretion model as a function 
of galaxy luminosity. Our final conclusions are summarized in Sect.\,\ref{Sect.Conclusions}.

%%%%%%%%%%%%%%%%%%%%%%%%%%%%%%%%%%%%%%%
%%%%%%%%%%%%%%%%%%%%%%%%%%%%%%%%%%%%%%%
\section{Data}\label{Sect:Data}

\subsection{HST/ACS imaging data}
The analysis of the specific frequencies of mostly late-type dwarf 
galaxies in this study is based on HST/ACS archival data from 
programs SNAP-9771 and GO-10235 (PI: I. Karachentsev), and 
GO-10210 (PI: B. Tully). These snapshot surveys provide 
Tip-of-the-Red-Giant-Branch distances to all dwarf galaxies in the 
studied sample \citep{Tully06, Karachentsev06, Karachentsev07}, 
and deep observations to study the GC systems of 
the dwarf galaxies. Our sample of dwarf galaxies consists of 55 dIrrs, 
3 dEs, 5 dSphs, and 5 Sm galaxies in the field environment, either 
in loose associations of dwarfs or in remote halo regions of nearby 
groups (Sculptor, Maffei 1\,\&\,2, IC 342, M 81, CVn I cloud). We 
have detected old GCs in 38 dwarf galaxies \cite[cf. Table\,1 in][]{Georgiev09}. 
Most of the dwarfs have apparent diameters smaller than the HST/ACS 
field of view which provides us with a good sampling of their 
GCSs. The GC numbers of two dwarf Sm galaxies from GO-10210 (NGC\,784 
and ESO\,154-023) and three Sm galaxies from SNAP-9771 and
GO-10235 (ESO\,274-01, NGC\,247, and NGC\,4605) 
have been corrected for spatial incompleteness. Details on the data 
reduction, cluster selection and photometry, completeness and total 
galaxy luminosity are described in detail in \cite{Georgiev08, Georgiev09}. 
\begin{figure*}
\includegraphics[width=1\textwidth, bb=25 10 740 220]{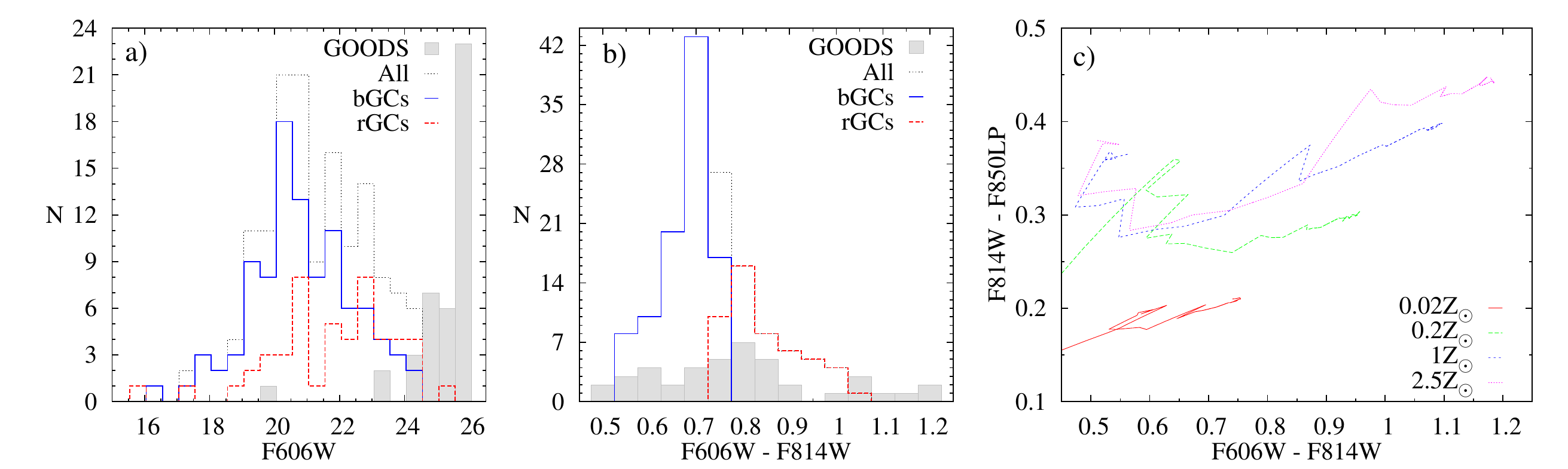}
\caption{Contamination analysis. {\bf Panels a)} and {\bf b)} 
show a comparison between the apparent luminosity and color distributions 
for GC candidates in our sample (all, blue and red GC samples with 
dotted, solid and dashed histograms) and those for sources in the 
GOODS survey (filled histograms) selected similarly to our initial 
GC candidates selection. Applying identical criteria for the selection 
of contaminants from the GOODS catalog was not possible due to the 
lack of F814W-band imaging in the latter. Magnitude and object 
morphological selection was based on the GOODS F606W-band measurements 
and an F606W$-$F850LP color cut derived from a comparison with SSP 
models, to include the entire range in metallicity at ages $>\!1$Gyr, 
matching the F606W$-$F814W color selection. Transformation from F606W$-$F850LP 
to F606W$-$F814W color requires knowledge of the object metallicity. 
This is well seen in the {\bf panel c)} where we show the F606W$-$F814W 
versus F814W$-$F850LP Galev SSP model tracks \citep{Anders03} for a 
range of metallicities indicated in the legend. For an approximate 
comparison, we applied a shift of 0.25\,mag to the GOODS F606W$-$F850LP 
color distribution in panel {\bf b)}.
}
\label{Fig.GOODS_contam}
\end{figure*}

\subsection{Contamination Analysis with HUDF and GOODS Data}\label{Sect:Contamination}
An important aspect which can affect the statistics of our analysis, 
especially for the faintest dwarf galaxies with fewer detected GC 
candidates, is the issue of background contamination. In \cite{Georgiev08} 
we used the ACS Hubble Ultra Deep Field (HUDF) to select objects with 
the same magnitude and morphological selection criteria as our sample GCs. 
Our analysis showed that the depth and resolution of the ACS images 
allowed us to exclude background contamination among the blue GC 
candidates, even at the previously adopted redder cut between blue 
and red objects at $(V-I)=1.1$\,mag. In our latest studies we have 
adopted a bluer cutoff between blue and red GC candidates at 
$(V\!-\!I)_0=1.0$\,mag which is the selection for the current study 
as well. For a more conservative analysis and for further narrowing 
the possibility of contamination we selected for this study GC candidates 
in the color range $0.7\!<\!V\!-\!I\!<\!1.4$\,mag, in accordance with 
the colors of Galactic and old LMC GCs.

In order to double-check the contamination level and avoid selection 
biases due to cosmic variance, we have accessed the latest version v2.0 
of the GOODS catalog \cite[Great Observatories Origins Deep Survey,]
[]{Gavalisco04}. GOODS is a deep imaging survey with HST/ACS in four 
bands ($B_{\rm F435W},V_{\rm F606W},i_{\rm F775W},z_{\rm F850LP}$) of 
two locations in the north and south sky covering a total of 
$\sim\!320$\,arcmin$^2$. We matched all sources with photometry in F606W 
and F850LP with a tolerance of 1\,pixel for both GOODS-N and GOODS-S 
catalogs. This yields a total number of 39\,432 and 33\,955 sources, 
respectively.

We analyze the contamination in a similar manner as in \cite{Georgiev08} 
by employing morphological information from the GOODS catalog and 
selecting sources with $19\!<\!{\rm F606W}\!<\!26$\,mag, 
$2\!<\!{\rm FWHM}_{\rm F606W}\!<\!9$ and ellipticity in the range 
$0\!-\!0.15$, identical to our F606W images. By applying the same 
objects selection strategy as for our science images on the entire 
high-level science GOODS v2.0 imaging data\footnote{Retrieved from 
http://archive.stsci.edu/prepds/goods/ and piped through the same 
analysis routines of F606W images.}
we confirmed the morphological selection. To constrain the selection 
of GOODS sources in a similar $V\!-\!I$ color range, as the selection 
of our GC candidates, we used the F606W$\!-\!$F850LP color. Observations 
of other galaxies, Galactic and Magellanic Cloud Clusters have 
Johnson/Cousins colors in the range $0.7\!<\!V\!-\!I\!<\!1.4$\,mag, 
which we convert using the Galev SSP models \citep{Anders03} to 
$0.74\!<\!$~F606W$-$F850LP~$\!<\!1.55$ for ages older than $\gtrsim
\!1$\,Gyr for the entire metallicity range. 
As a result, 18 and 24 objects passed the above selection criteria for 
the North and South GOODS fields, respectively. Combined and normalized 
to the total GOODS survey area of $320$\,arcmin$^2$ leads to an upper 
limit of the total contamination density of $\rho\!\approx\!0.13$ objects 
per arcmin$^2$, i.e. $\sim\!1.5$ objects in a single ACS field with an 
area of 11.33\,arcmin$^2$. In total we have 30 galaxies with blue GC 
detections (see Sect.\,\ref{subSectSN}), i.e.~$\sim\!340$\,arcmin$^2$ 
which is comparable to the total area of the GOODS fields. Thus, virtually 
no scaling to the number of contaminating objects is required for the 
direct comparison shown in Figure\,\ref{Fig.GOODS_contam}. The luminosity 
and color distributions of all contaminating sources are shown in panels 
a) and b). The magnitude difference between F814W and F850LP shown in 
panel c) is in the range $0.15\!-\!0.45$\,mag. This shows that without 
knowing the object metallicity it is not possible 
to derive precise information about the contamination as a function of 
color. To partially account for this difference and ease the comparison 
we shifted the color distribution of GOODS objects in Figure\,\ref{Fig.GOODS_contam}\,b) 
by 0.25\,mag. 
The luminosity distribution of background objects reveals that the 
contamination in our entire GCs sample is of the order of $\sim\!5\!-\!6$ 
among the faintest, barely resolved GC candidates. 
This additional analysis of two different sky regions with deep ACS 
imaging confirms the expected low level of contamination. Section\,\ref{Sect:S_N_for_our_sample} 
discusses the implications of this result.

%%%%%%%%%%%%%%%%%%%%%%%%%%%%%%%%%%%%%%%
%%%%%%%%%%%%%%%%%%%%%%%%%%%%%%%%%%%%%%%
\section{Analysis}\label{Sect:Analysis}

%%%%%%%%%%%%%%%%%%%%%%%%%%%%%%%%%%%%%%%
\subsection{Specific Frequency}\label{subSectSN}
The globular cluster specific frequency is defined as 
\begin{equation}
S_N=N_{\rm GC}\times10^{0.4(M_V+15)} \label{eqn:sn}
\end{equation}
\citep{Harris&vdBergh81} and is a measure of the ratio of the formation
efficiencies of star clusters to field stars. It describes how likely 
it is for a given galaxy to have formed star clusters that survived 
the various cluster disruption mechanisms until today. Two factors affect 
the $S_N$-values in late-type dIrr galaxies: the recent star 
formation activity of the host galaxy affects $M_V$ while the dynamical 
evolution history of its GC system affects $N_{\rm GC}$.

In particular, for our sample galaxies the influence of changes in 
galaxy luminosity is more significant for late-type dwarf galaxies 
for which the present-day starburst activity can boost their luminosity, 
causing a decrease in the $S_N$ values. Formation and subsequent 
dynamical evolution of GCs in low-mass galaxies (where fewer clusters 
are formed) leads to stochastic fluctuations in $N_{\rm GC}$ which
can lead to an increased dispersion of $S_N$ among such dwarfs, which 
is indeed observed. In addition, the stochastic nature of the field-star 
formation history for dIrrs \citep[e.g.][]{Weisz08} can introduce additional 
scatter. 

In spite of this, the field star and star-cluster formation efficiencies
generally appear to be closely correlated as it is observed that the ratio of
cluster to  star formation rates (SFR) is nearly constant \cite[e.g.][]{Lada&Lada03, 
Lamers&Gieles08, Gieles&Bastian08} and
independent of the present-day SFR \citep{Bastian08}. In addition, the
constant ratio between star-cluster mass and the mass in field stars at
the end of violent relaxation\footnote{The end of violent relaxation, when
the dynamical response of a cluster to the expulsion of its leftover star
forming gas is completed, typically takes place at an age between 10 and
50\,Myr and depends on the external tidal field and the proto-cluster
crossing-time \protect\cite[see Fig.\,4 in][]{Parmentier09}.}~(i.e.~the bound
cluster-to-field stellar mass ratio) may be evidence for a nearly
universal distribution function of the {\it local} star formation efficiency
\citep{Parmentier&Fritze09}.

\begin{figure}
\includegraphics[width=0.5\textwidth, bb=25 25 360 260]{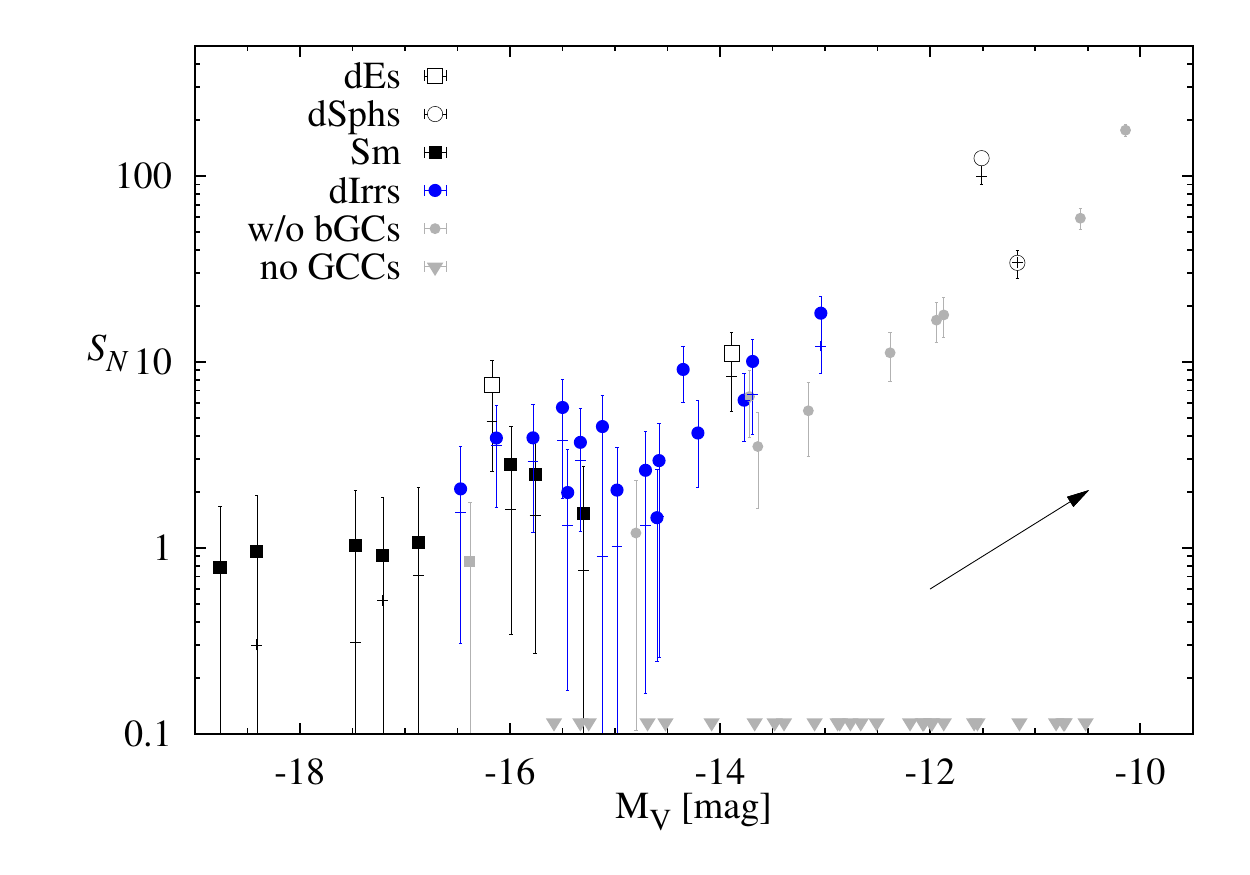}
\caption{Globular cluster specific frequency, $S_N$, versus absolute 
galaxy magnitude, $M_V$, for all dwarf galaxies in our sample. Symbols 
are the bGC$+$rGC estimates of the $S_N$ values for various galaxy 
types, while plus signs indicate the values based on the bGC population 
only.Grey triangles at the bottom of the plot represent galaxies in 
which no GCCs were detected.
The black arrow indicates the change of the $S_N$ value due to passive
evolutionary fading of the galaxy light by $\Delta M_V=1.5$\,mag (from 3 to
14\,Gyr at LMC metallicity). 
\label{Fig.SN}}
\end{figure}
\subsubsection{Specific frequencies of dwarf galaxies in our sample}\label{Sect:S_N_for_our_sample}
We compute two $S_N$ values for our sample of GC candidates selected 
from \cite{Georgiev08,Georgiev09} in the color range $0.7\!<\!V-I\!<\!1.4$\,mag. 
First using the number of blue GCs and second the total number of GCs: blue 
GCs\,$+$\,red GCs (red GCs have $(V\!-\!I)_0\!>\!1.0$ mag), together with 
the total $V$-band galaxy magnitudes derived in \cite{Georgiev08, Georgiev09}. 
The calculated $S_N$ values for all dwarf galaxies in our sample are 
shown in Figure~\ref{Fig.SN} and listed in Table~\ref{Table:SN}. 
The contamination analysis in Section\,\ref{Sect:Contamination} 
suggests that red GC candidates with $1.0\!<\!(V\!-\!I)\!<\!1.4$ can not be fully 
accounted for by contamination in our sample galaxies. This is puzzling 
because the existence of such red, metal-rich GCs in low-mass dwarf 
galaxies is not expected. The fraction of red GCs in early-type galaxies 
decreases with decreasing galaxy luminosity and becomes negligible for 
the faintest dEs \cite[cf Fig.\,4 and 10 in][]{Peng06}. Theory suggests 
that red, metal-rich clusters are believed to form either in major 
galaxy mergers \cite[e.g.][]{Ashman&Zepf92} or in a multi-phase collapse 
of a massive galaxy \citep{Forbes97,Pipino07}. Both channels are unlikely 
to occur in low-mass systems. Observationally, extinction within the 
galaxy can lead to a redder colour. Taking the Magellanic Clouds as a 
proxy for the dwarfs in our sample, the expected internal reddening 
is in the range $0\!<\!E(V\!-\!I)\!<\!0.13$\,mag 
\footnote{The total $E(B\!-\!V)$ value for old LMC GCs is in the range 
$0.083-0.213$\,mag for NGC\,1898 and NGC\,1841 \citep{McLaughlin&vdMarel05}. 
Corrected for the median Galactic foreground reddening toward LMC of 
$E(B-V)=0.075$\,mag \citep{Schlegel98}, leads to $E(B\!-\!V)=0.008$ 
and 0.138\,mag, i.e. $E(V\!-\!I)=0.009$ and 0.164\,mag internal reddening 
for the LMC. The same analysis for the only old SMC GC, NGC\,121, gives 
$E(B\!-\!V)=0.109$\,mag or $E(V\!-\!I)=0.130$\,mag. The LMC GCs above 
are at $R=0.6^\circ$ and $14.9^\circ$ projected distance from the center 
of the LMC \citep{vdBergh&Mackey04}. $m\!-\!M=1$8.5\,mag (50.118\,kpc) 
to the LMC yields 0.5 and 13.3\,kpc.}
The radial distribution of the bGCs and rGCs in our sample galaxies are 
indistinguishable from each other. They are distributed mostly within 
1-2\,pc projected distance from the galaxy center, i.e. $\sim1\times$R$_{\rm eff}$ 
\citep{Georgiev09b}. Therefore, some internal reddening can be expected 
among the GCs in our sample. Red colors could arise if clusters have ages 
in the range $1-3$\,Gyr where AGB stars are in their thermal-pulsation 
phase (TP-phase) which affects the integrated cluster color. Comparison 
with population synthesis models \cite[e.g.][]{Maraston05,Leitherer99} 
shows that for that age range the integrated cluster colour can reach 
$V\!-\!I\!<\!1.4$\,mag for even as low metallicities as $0.2$ to $0.4\,Z_\odot$. 
However, whether reddening or younger ages are affecting the colors of 
GCs in our sample, or whether they are background contaminants will 
require spectroscopic or near-IR follow-up observations to establish 
their accurate metallicities and membership.

In order to minimize the effects on the $S_N$ from the uncertain nature 
of the rGCs in our subsequent analysis, we consider \emph{only} dwarf 
galaxies which contain blue GC candidates. There are eight galaxies ($20\%$) 
with only red GCs which are treated as galaxies with null GC detections. 
The lack of bGCs does not necessarily mean that some of the rGCs cannot 
be real GCs due to the effects discussed above. Therefore, given these 
uncertainties, our $S_N$s represent a conservative {\it lower} value for 
faint late-type galaxies. In Section\,\ref{Sect.ContinuousRelations} we 
discuss the implications and limitations of the presence of rGCs for the 
model we are testing.

Overall, the observed trend of increasing $S_N$ with decreasing host galaxy 
luminosity holds for the entire sample, and in particular for dIrrs. 
Because of their small sample size the trends for dEs/dSphs and Sm 
galaxies are less conclusive from our data alone. Grey triangles 
show galaxies where no GC candidates are detected. Their $S_N$ values 
were adjusted for illustrative purposes only. Notably, Figure\,\ref{Fig.SN} 
shows that for low-luminosity galaxies the difference between total 
$S_N$ (where the rGCs sample can include background contaminants) and 
blue GC $S_N$ (the most trustworthy value) differ a little quantitatively, 
but not significantly in the qualitative sense due to the logarithmic 
scaling. Thus, the observed general trend of increasing $S_N$ value 
with decreasing galaxy luminosity is present even if only $S_N$ values 
calculated from the blue GCs (bGCs) are considered.

It is important to note that all $S_N$ values for the late-type dwarfs 
represent their present-day values. Passive evolutionary fading of the 
integrated galaxy light will increase the $S_{N}$ values by a factor 
of 2 to 16 for our dIrr and Sm galaxy sample (assuming $N_{\rm GC}=$\,const.).
This is shown by an arrow in Figure~\ref{Fig.SN}, which indicates the 
change in $S_N$ if a galaxy passively fades due to stellar evolution by
$\Delta M_V=1.5$\,mag. This corresponds to an evolution from 3 to 14 Gyr
at LMC metallicity as inferred from SSP models \cite[e.g.][]{BC03}.

\subsubsection{Comparison of $S_N$ values of late-type dwarf galaxies}\label{Sect:S_N_for_our_sample_compared_with_liter}
\begin{figure}
\includegraphics[width=0.5\textwidth, bb=20 25 360 260]{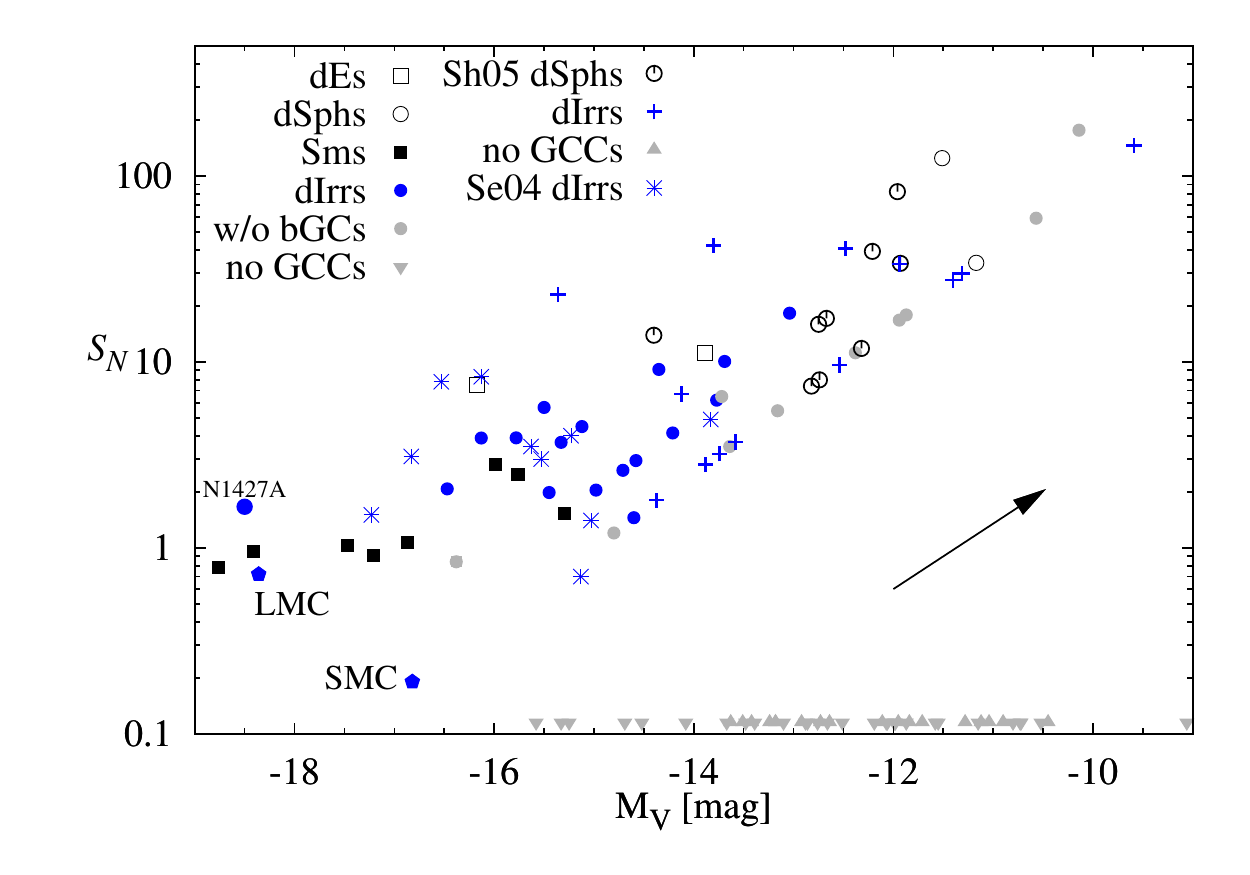}
\caption{GC specific frequency as a function of absolute galaxy magnitude 
for dwarf galaxies at low density environment from 
\protect\cite{Sharina05,Georgiev08,Georgiev09} and dIrrs in Virgo and
Fornax cluster \protect\cite{Seth04,Georgiev06}. Dwarf irregular galaxies
are shown with solid circles, plus signs and asterisks; dSphs with open
and dashed-open circles; Sms with solid squares. With solid pentagons we
show $S_N$ values of the best studied GC systems in nearby dIrrs --
the LMC and SMC. Grey triangles at  the bottom of the plot  represent 
galaxies in which no GCCs were detected. 
The arrow indicates passive evolutionary fading as in Figure~\ref{Fig.SN}.
\label{Fig.SNcompare}}
\end{figure}
In Figure~\ref{Fig.SNcompare} we compare the $S_N$ values of our sample 
dwarfs with those of late-type dwarfs from \cite{Sharina05}, all in 
field environment, and Virgo and Fornax cluster dIrrs from \cite{Seth04}  
and \cite{Georgiev06}, as well as the Magellanic Clouds. It is clearly seen
that, combined with the data from the literature, we observe a general
trend of increasing $S_N$ with decreasing galaxy luminosity for dIrrs. 
Such a trend has been previously observed for early-type dEs \cite[e.g.][]{
Miller98,Strader06}. The dIrrs in our sample are systematically brighter 
than dSphs by $\sim 3$ mag, indicating younger ages. As mentioned above, 
a simple picture of passive evolutionary fading of the dIrrs can shift their 
$S_N$ values toward the dSphs region, parallel to the average trend. 
This supports the notion that if the number of GCs remains unchanged the 
high$-z$ analogs of present-day dIrrs might be the progenitors of the more 
evolved dSphs/dEs \cite[e.g.][]{Miller98, Seth04}.

\begin{figure}
\includegraphics[width=0.5\textwidth, bb=20 25 550 550]{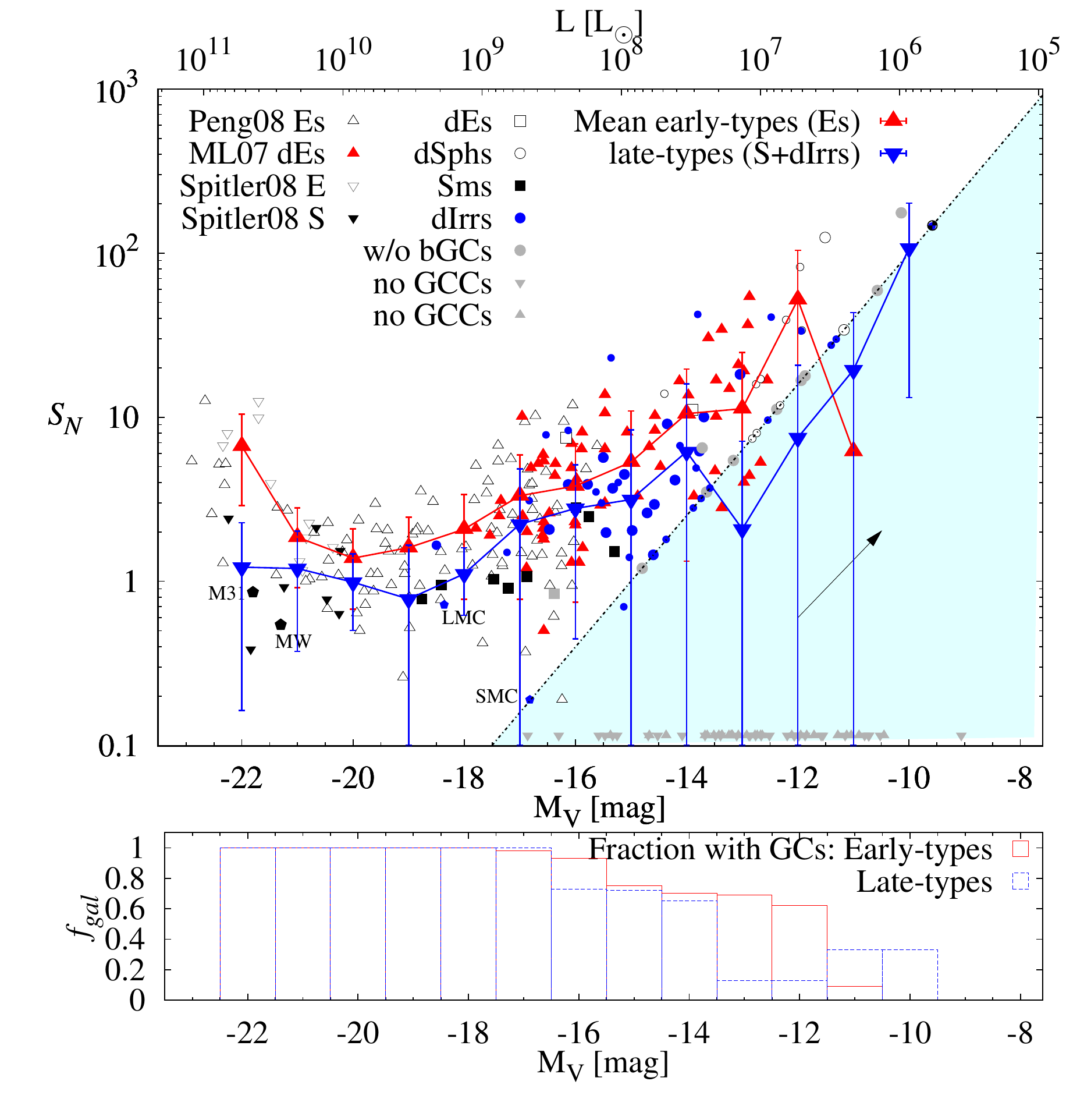}
\caption{{\bf Top:} GC specific frequency versus absolute galaxy magnitude 
for a range of galaxy masses, morphologies, and environments. The various 
symbol types are explained in the figure legend. Massive and dwarf elliptical
galaxies as well as spirals have been collected from the literature as 
explained in Sect.\,\ref{subSectSN}. With larger symbol sizes (dots, 
circles and squares) we indicate the dwarf galaxies in our sample. Grey 
triangles at the bottom of the plot represent galaxies in which no GCCs 
were detected. The large symbols connected with a line are the co-added 
running average of the $S_N$ values per magnitude bin for the different 
galaxy morphological types. The errorbars are the standard deviation at 
that magnitude bin. The dash-dotted line indicates the $S_N$ value if a 
galaxy hosts one GC (cf. Eqn.\,\ref{eqn:ll}). {\bf Bottom:} Fraction of 
galaxies with GC detections for the magnitude bins used to derive the 
mean $S_N$ value in the upper panel.
\label{Fig.SNcompare2}}
\end{figure}
To investigate the formation efficiency of GCs as a function of galaxy 
luminosity, we extend the galaxy mass range and morphology by including  
data from the literature. In Figure\,\ref{Fig.SNcompare2} we compare the
specific frequencies of the dwarf galaxies discussed above with those of
early-type galaxies in the Virgo galaxy 
cluster from \cite{Peng08} and other massive elliptical and spiral galaxies 
from the compilation by \cite{Spitler08} who selected galaxies with robustly 
derived GCS properties (deep photometry and completeness limits, adequate
spatial coverage etc.) from the literature \citep{Ashman&Zepf98,Rhode&Zepf01,
Rhode&Zepf03,Rhode&Zepf04,Forbes01,Forte01,Dirsch03,Dirsch05,
Gomez&Richtler04,Bassino06a,Bassino06b,Bassino08,Harris06a,Tamura06a,Tamura06b,Rhode07}.   
Our main source for the early-type dwarf galaxy data is \cite{Miller&Lotz07}. 
For reference, we have labeled in Fig.\,\ref{Fig.SNcompare2} the corresponding 
$S_N$ values of the Magellanic Clouds, M31, and the Milky Way. 
Keeping in mind the effects of passive evolutionary fading and star 
cluster disruption, the $S_N$ vs. $M_V$ trends for late- and early-type 
dwarfs in cluster and in loose environments appear very similar irrespective 
of galaxy morphological type and environment. 
With a dash-dotted line we indicate the lower limit of the $S_N$ value
defined by Equation~\ref{eqn:sn}  if the galaxy hosts one GC ($N_{\rm GC}=1$).

To investigate the $S_N$ value as a function of galaxy morphology 
and luminosity we have estimated the mean $S_N$ value in bins of 
one magnitude. 
With large open symbols connected with lines in Figure\,\ref{Fig.SNcompare2} 
we show these co-added running averages of the $S_N$ including 
galaxies without GC candidates. They have been computed by summing 
the number of GCs and galaxy luminosity for each magnitude bin. 
The error bars represent the standard deviation at that magnitude 
bin. It can be seen that late-type galaxies (spirals and irregulars) 
have on average ($\sim2$ times) lower mean $S_N$ values at a given
galaxy luminosity than early-type galaxies (ellipticals and spheroidals). 
This constant offset can be accounted for by evolutionary fading 
effects. The spiral galaxies in our sample show a fairly constant 
$S_N$ value as a function of galaxy luminosity ($S_N\!\approx\!1$). 
However, a larger sample increasing the number of spiral galaxies 
with luminosities $M_V\!<\!-19$\,mag need to be studied to confirm the 
nearly constant $S_N-$value for spiral galaxies.

The bottom panel of Figure\,\ref{Fig.SNcompare2} shows an increasing 
fraction of galaxies with no detected old GCs towards decreasing galaxy 
luminosities. This decrease can be attributed to several factors affecting 
the galaxy luminosity normalized number of GCs: $i)$ The increased 
impact of stochasticity for star-cluster formation and/or $ii)$ the 
decrease of GC numbers due to dynamical destruction processes (i.e. 
evaporation, tidal disruption, etc.).

%%%%%%%%%%%%%%%%%%%%%%%%%%%%%%%%%%%%%%%
\subsection{Specific Mass, Luminosity, and $\hat{T}$-value}\label{SmSect}
In addition to the traditionally used specific frequency $S_N$, we discuss in
the following the trends in the specific mass and specific luminosity
of the GC system, 
as well as a redefined $T$-value \citep[see][]{Zepf&Ashman93} as a function of
host galaxy luminosity and mass. 

\subsubsection{Specific Mass of GC system}

We define the ``specific mass'' $S_{M}$ as the total mass of the globular 
cluster system, ${\cal M}_{\rm GCS}$, relative to the total baryon mass of the host
galaxy defined as the sum of the stellar mass, ${\cal M}_\star$, and 
the H{\sc I} gas mass, ${\cal M}_{\rm H{\sc I}}$:
\begin{equation}\label{eqn:sm}
S_{M}=100\times {\cal M}_{\rm GCS}/({\cal M}_\star+{\cal M}_{\rm H{\sc I}}).
\end{equation}
This is similar to 100 times the \cite{McLaughlin99} $\epsilon$ parameter
for the GC formation efficiency for giant galaxies with respect to the 
total baryon mass, which includes the mass of the hot, X-ray emitting
halo. It is not a trivial task to include this mass component for the galaxies 
in our sample. Obtaining hot gas mass from X-ray profiles ($M_X$) relies on 
assumptions of the temperature profile as well as hydrostatic equilibrium, which 
is not always the case for galaxies in complex environments. For instance, 
the \cite{Fukazawa06} X-ray observations of luminous early-type ellipticals in a range 
of environments show much weaker X-ray emission from isolated 
than for clustered galaxies. In addition, they find that the requirement for hydrostatic 
equilibrium might not be fulfilled for many of their ellipticals due to jets from AGNs 
and interaction with neighboring galaxies. In general, all cD galaxies in their sample 
(like NGC\,1399) are classified as galaxies embedded in the hot cluster/group ISM. 
Therefore, the X-ray halo of such central cluster galaxies may have a non-negligible 
contribution from the galaxy cluster itself. 
To avoid 
problems with hydrostatic equilibrium \cite{Humphrey06} and \cite{O'Sullivan07} 
studied isolated elliptical galaxies for which they obtain $M_X$ few times $10^8M_\odot$, 
a similar $M_X$ mass was estimated for M\,49 by \cite{McLaughlin99}. 
Therefore, for the majority of the galaxies in our sample the expected 
$M_X$ is of the order of few percent of the stellar mass, thus a negligible 
contribution to the total galaxy mass.

In the following we have considered the sum of the H{\sc I} and stellar
masses, ${\cal M}_{\rm\star}+{\cal M}_{H{\sc I}}$, for the dwarf galaxies in
our sample. Including the H{\sc I} gas mass becomes more important when comparing 
massive elliptical galaxies with dwarf galaxies which can contain significant 
neutral gas mass fractions.	The disadvantage of the $S_M$ parameter (Eq.~\ref{eqn:sm}) is the
necessary conversion from luminosity to mass, using ${\cal M}/L$
ratio for the galaxy as well as the GCs (see below). On the other hand, $S_M$
has the advantage of being a distance-independent quantity. 

To derive ${\cal M}_{\rm GCS}$ for dIrrs we convert GC absolute
magnitudes to masses using $\gamma_V\!\equiv\!({\cal M}/L_V)_{\rm
GC}\!=\!1.88$, which is the mean $V$-band mass-to-light ratio for old GCs
in the LMC that we estimated from \cite{McLaughlin&vdMarel05}. Since we
are studying old globular clusters
in similar galaxies, our choice of $\gamma_V$ is justified and changes only
within 10\% for the age range of our GC sample selection ($t\!\ga\!4$
Gyr). However, a recent study by  \citet{Kruijssen&Zwart09} showed that the GC
${\cal M}/L$ ratio is expected to depend on cluster luminosity (mass) in that  
low-mass clusters preferentially lose low-mass, i.e. high ${\cal M}/L$, 
stars due to two-body relaxation. They show that this effect influences the 
translation from GC luminosity to mass function and the respective 
location of the turn-over mass ($m_{\rm TO}$) toward lower mass by 
$\sim0.2$\,dex. This effect needs to be tested for a range of galaxy 
masses (i.e., tidal field strengths) in order to establish the GC ${\cal M}/L
\propto L$ effect  to the GCS $m_{\rm TO}$. Keeping this in mind and assuming
a constant GC ${\cal M}/L$ for the GCSs of all galaxies, we will show in
Sect.\,\ref{Sect:Discussion} that this effect is 
negligible relative to the observed spread/scatter in the relations that
involve the GC ${\cal M}/L$ ratio. 

\begin{figure}
\includegraphics[width=0.5\textwidth, bb= 15 25 360 260]{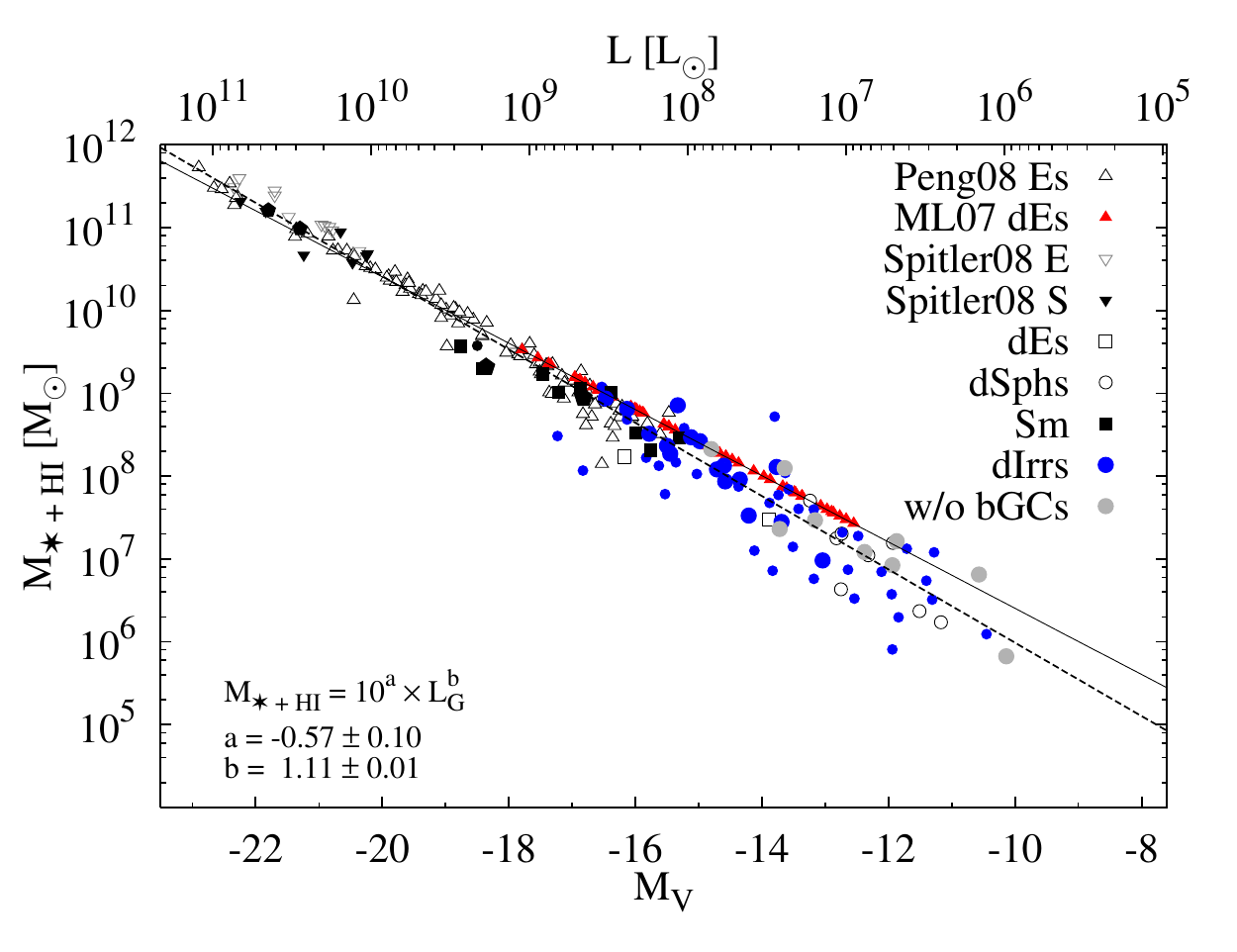}
\caption{Stellar plus H{\sc I} mass as a function of galaxy luminosity 
for dwarf galaxies in our sample (large solid dots) compared with galaxies of
different morphological type from the literature as indicated in the
legend. Smaller solid dots show dIrrs from \protect\cite{Seth04} and
\protect\cite{Sharina05}. With solid pentagons are shown (frm left to right)
M\,31, Milky Way, LMC and SMC. The dashed line is a least-squares fit to the data; the fit
relation and the values of the coefficients are given in the lower left
corner. The solid line shows the relation between the galaxy luminosity
and its mass with ${\cal M}_\star/L_V=3.0$.
\label{Fig:Mt_L}}
\end{figure}

${\cal M}_{\rm H{\sc I}}$ was calculated from the H{\sc I} magnitudes listed 
in the LEDA database \citep{Paturel03} or from H{\sc I} fluxes in NED or \cite{Begum08}  
following \cite{Roberts&Haynes94}, i.e. $M_{\rm H{\sc I}}=2.36\times10^5
\times D^2\times F_{\rm H{\sc I}}$, where $D$ is the distance in Mpc and 
$F_{\rm H{\sc I}}$ is the H{\sc I} flux in Jy km\,s$^{-1}$. To convert from 
galaxy luminosity to galaxy stellar mass ${\cal M}_\star$, we used the 
relations in \cite{Bell03} between the galaxy $B\!-\!V$ colour and the 
${\cal M}_\star/L_V$ ratio (or $B\!-\!R$ if $B\!-\!V$ was not available):
\begin{eqnarray*}
\hspace*{1.05cm}\log({\cal M}_\star/L_V)=-0.628+1.305\times(B\!-\!V)\hspace*{0.08cm}\\
\hspace*{1.05cm}\log({\cal M}_\star/L_V)=-0.633+0.816\times(B\!-\!R).
\end{eqnarray*}
In addition to our dIrr galaxy data, we calculated ${\cal
M}_\star/L_V$ ratios for elliptical and spiral galaxies from \cite{Peng08,
Spitler08, Miller&Lotz07} and for dSphs and dIrrs from \cite{Sharina05}
and \cite{Seth04} as follows. We use the stellar masses directly from the
\citeauthor{Peng08} study which were computed from galaxy colours
($g\!-\!z$ and $J\!-\!K_s$). \cite{Spitler08} used $K-$band photometry and
theoretical ${\cal M}_\star/L_K$ from population synthesis models to derive 
galaxy mass, which we also adopt in the subsequent analysis. To derive
baryonic masses for the dE  
sample in \cite{Miller&Lotz07} we use the galaxy magnitudes provided in their 
table and ${\cal M}_\star/L_V=3.0$. We used this mass-to-light ratio because 
the masses listed in their table were calculated assuming ${\cal
  M}_\star/L_V=5$ which seems to be too high and offset from the
relation defined by our dataset and those of \cite{Seth04,Sharina05,
  Peng08,Spitler08} (see Fig.~\ref{Fig:Mt_L}). For the dwarf galaxies  
in the \cite{Sharina05} and \cite{Seth04} samples we calculate their masses 
from the \citeauthor{Bell03} relations as we did for our sample. 
Note that uncertainties in the galaxy ${\cal M}/L$ determination due to
unknown SFHs and the choice of an IMF apply equally to
the range of galaxy masses in our sample as to that of the SDSS and 2MASS
galaxy samples from which the  \citeauthor{Bell03} relations were derived. 

In Figure~\ref{Fig:Mt_L} we illustrate the relation between galaxy
luminosity and its stellar plus H{\sc I} mass, ${\cal M}_{\rm\star+H{\sc I}}$,
for galaxies from the literature and the dwarf galaxies in our sample. 
We find an increasing difference with decreasing luminosity between 
the late-type, gas-rich, low-mass dIrrs, 
and early-type dwarf ellipticals, which are essentially devoid of H{\sc I}. 
In addition, the low-mass dIrrs show an increasing dispersion towards 
fainter systems. This offset and dispersion can both be attributed to 
possibly underestimated ${\cal M}_b/L_{V}$ due to younger stellar populations
and/or unaccounted-for baryonic 
components in the form of, e.g.~diffuse ionized gas, hot X-ray emitting
gas, radio continuum emission or molecular gas. All these mass components are 
known to be associated with star formation activity and
their mass fractions are observed to scale with galaxy H{\sc I} mass
\cite[e.g.][]{Dahlem06}. 

\subsubsection{Specific Luminosity of GC system}\label{Sect:SL}

Another useful indicator of GC formation efficiency is the specific GC 
luminosity introduced by \cite{Harris91}
\begin{equation}
S_{L,V} = 100\times L_{\rm GCS}/L_V
\label{eqn:sl}
\end{equation}
which is the luminosity ratio between the GCS, $L_{\rm GCS}$, and its host
galaxy, $L_V$. Its main advantage as scaling parameter is that it is
independent of distance and only weakly sensitive to completeness corrections
at the faint end of the GC luminosity function (GCLF). A slight disadvantage
of $S_L$ is its sensitivity to the stellar population dominating the integrated
light of the galaxy. For the case of late-type galaxies this typically is the
most prominent (the latest) star formation episode, so that the galaxy
luminosity does not necessarily indicates the stellar population at the time when 
most of the GCs were formed. On the other hand, $S_L$ is independent of ${\cal M}/L$ and hence 
least affected by stellar population systematics. Thus, $S_L$ may be
the most robust GCS scaling parameter for comparing GCS richness and GC 
formation efficiencies between galaxies over large range of masses.

\subsubsection{Modified $T$-value}

\cite{Zepf&Ashman93} introduced the $T$ value as the specific number of
globular clusters per unit $10^9M_\odot$ of the host galaxy mass,
i.e. $T=10^9M_\odot\times N_{\rm GC}/{\cal M}_\star$. For our purposes this
definition is insufficient 
since we want to compare gas-depleted early-type with late-type galaxies
with significant gas mass fractions. Thus, we modify the $T$-value from 
\citeauthor{Zepf&Ashman93} by including the gas mass of the host galaxy. 
Hence we define the specific number of GCs per unit baryonic mass $\hat{T}$ 
(${\cal M}_{b}$) as follows:
\begin{equation}
\hat{T}=10^9M_\odot\times N_{\rm GC}/{\cal M}_{b}
\label{eqn:T}
\end{equation}
where ${\cal M}_{b}\!=\!{\cal M}_\star\!+\!{\cal M}_{\rm H{\sc I}}$.

In Figure~\ref{Fig:SNLMT_compare} we directly compare the GC specific
frequencies ($S_N$, Eq.\,\ref{eqn:sn}), the mass-normalized specific mass
in clusters ($S_M$, Eq.\,\ref{eqn:sm}), specific luminosities ($S_L$,
Eq.\,\ref{eqn:sl}), and the specific number ($\hat{T}$, Eq.\,\ref{eqn:T})
as a function of host galaxy luminosity for the entire sample.

%%%%%%%%%%%%%%%%%%%%%%%%%%%%%%%%%%%%%%%
%%%%%%%%%%%%%%%%%%%%%%%%%%%%%%%%%%%%%%%
\section{Discussion}\label{Sect:Discussion}

In this section we address the question whether there is a physically motivated 
model which is able to describe the observed trends of the various scaling 
relations as a function of total galaxy mass. This can be tested by assuming 
that GCs form in proportion to the total galaxy mass. We start by comparing 
the GC scaling relations introduced in the previous Section\,\ref{Sect:Analysis} 
with theoretical model predictions that provide relations between the
mass-to-light ratio and the total galaxy mass \citep{Dekel&Silk86, 
Dekel&Birnboim06}. The star-formation processes in these models are driven 
by {\it (i)} stellar feedback and {\it (ii)} shock heating below and above 
a critical stellar mass of $3\times10^{10}{\cal M}_\odot$, respectively. 
Thus, we derive analytical expressions that can be used to confront the 
observed trends  against the assumptions of constant {\it and} variable GC
formation efficiencies as a function of galaxy luminosity. Observed 
differences and systematics can argue for or against physical mechanisms 
included (or the lack thereof) in the models we discuss.

Is the assumption of an universal mode of globular cluster formation 
among galaxies of different morphological type and environment justified? 
An universal GC formation scenario would predict that all (evolved) 
galaxies form their oldest star cluster populations with the same 
initial efficiency at given galaxy mass. Subsequent formation of 
metal-rich GCs in massive galaxies would likely cause a spread in the 
observed GC formation efficiency value for a given range in galaxy
mass since galaxies undergo different merging histories, varying from
predominantly dissipational to mostly dissipationless. This can be 
tested by comparison of the expected value of the GC efficiency from 
the adopted analytical models and the observations of galaxies of 
the two major morphological types, early- versus late-types,
and as a function of galaxy mass. The relative difference between 
the two galaxy types should be preserved at a given galaxy mass, 
assuming that GC destruction processes are similar to first order 
in both galaxy types.

In order to set quantitative constraints and 
limitations of the models we use, and address all questions discussed 
above, it is very important to normalize the model predictions, 
i.e. to put the models on an absolute scale. All studies, so far, which 
addressed the shape of the $S_N$ distribution used one or another version of the
\citeauthor{Dekel&Birnboim06} models \cite[e.g.][]{Forbes05} but did 
not derive strict constraints due to the lack of such a normalization. 
Therefore, our effort in deriving a normalization will help to narrow 
down the possible physical processes that govern the spread in GC 
formation efficiencies among galaxies with similar mass. This is 
detailed in Section\,\ref{Sect:GCscalingNorm}.

%%%%%%%%%%%%%%%%%%%%%%%%%%%%%%%%%%%%%%%
\subsection{Boundaries of Globular Cluster Systems Scaling
  Relations}\label{Sect:LowerLimit} 
The definition of $S_N$ (Eq.\,\ref{eqn:sn}) determines the most trivial 
lower limit of star cluster population when $N_{\rm GC}\!=\!1$, i.e. the 
GC formation efficiency that resulted in only one old GC at present
(and no additional field star component), so that
\begin{equation}
\log S_N=6+0.4 M_V =7.928-\log L_V , \label{eqn:ll}
\end{equation}
assuming $M_{V,\odot}=4.82$ mag \cite[]{Cox00}. Analogously, we can 
derive similar relations for  $S_L, S_M$ and $\hat{T}$ by assuming 
that the total luminosity and mass of the globular cluster system are 
$L_{\rm GCS}=N_{\rm GC}\times l_{\rm TO}$ and ${\cal M}_{\rm GCS}=
N_{\rm GC}\times m_{\rm TO}$, where $l_{\rm TO}$ and $m_{\rm TO}$ are 
the cluster luminosity and cluster mass at the turn-over of the GCLF, 
respectively, hence
\begin{eqnarray*}
\log S_L=2+\log l_{\rm TO}-\log L_V
\end{eqnarray*}
\begin{equation}
\hspace*{0.975cm}=2+\log l_{\rm TO}+0.4\,(M_V-M_{V,\odot}) \label{eqn:Sll}
\end{equation}
\begin{eqnarray*}
\log S_M = 2 + \log(m_{\rm TO}/\Upsilon_V) - \log L_V
\end{eqnarray*}
\begin{equation}
\hspace*{1.05cm}=2 + \log(m_{\rm TO}/\Upsilon_V) + 0.4\,(M_V-M_{V,\odot}),\label{eqn:Sml}
\end{equation}
\begin{eqnarray*}
\log\hat{T}=9-\log\Upsilon_V-\log L_V
\end{eqnarray*}
\begin{equation}
\hspace*{0.81cm}=9-\log\Upsilon_V+0.4\,(M_V-M_{V,\odot}) \label{eqn:Tl}
\end{equation}
where $\Upsilon_V\equiv {\cal M}_\star/L_V$ is the (overall) $V$-band stellar 
mass-to-light ratio of the host galaxy. The relations (\ref{eqn:ll}) to
(\ref{eqn:Tl}) are shown in Figure~\ref{Fig:SNLMT_compare} with 
dash-dotted lines. For the derivation of $S_M$ and $\hat{T}$ we have adopted
$\Upsilon_V=3.0$, $l_{\rm TO}=8.47\times10^{4}L_\odot$, and $m_{\rm
TO}=1.69\times10^{5}M_\odot$, which corresponds to $M_V=-7.5$\,mag
for $\gamma_V\!=\!2.0$. 
Note that a change from $\Upsilon_V\!=\!3.0$ to 1.0 and
from $M_V=-7.5$ to $-7.4$\,mag causes a parallel shift comparable to the
symbol size relative to the 
relations.

Since the quantities $S_N$, $S_L$, $S_M$, and $\hat{T}$ involve properties of
GC systems that are (often) measured down to the turnover luminosity of the GC
luminosity function, it is useful to assess the uncertainties associated
with Eqs.\ \ref{eqn:ll} -- \ref{eqn:Tl}. 
The results of \cite{Jordan07} show that the dispersion $(\sigma)$ and 
turn-over luminosity $(l_{\rm TO})$ of the GCLF changes as a function of 
host galaxy luminosity. The same study shows also that the mass at the GCLF
turnover is roughly  
constant \cite[see also][]{Vesperini98, Vesperini00} and decreases by $\sim30\%$ 
from $m_{\rm TO}=(2.2\pm0.4)\times10^5 M_\odot$ in the brightest galaxies 
down to $m_{\rm TO}\simeq(1.6\!-\!1.7)\times10^5 M_\odot$ in dwarf elliptical
galaxies, with a substantial scatter in the latter. For comparison, in our analysis 
of the GCLF in mainly late-type dwarf galaxies we derive a GCLF turnover mass
$m_{\rm TO}=1.6\times10^5 M_\odot$ \citep{Georgiev08} assuming
$\gamma_V=1.88$, which is the mean value of the mass-to-light ratio of old
GCs in the LMC and SMC dIrrs \citep{McLaughlin&vdMarel05}. Due to the fact that 
the HST/ACS field of view entirely covers the dwarf galaxies in our sample we observe 
the total GC population and derive an accurate {\it total} $S_N$\footnote{Note that the
original definition of \cite{Harris&vdBergh81} is to double the number of
GCs brighter than the GCLF turnover luminosity to obtain $N_{\rm GC}$ in
Equation~\ref{eqn:sn}. This is identical to the total $N_{\rm GC}$ only if the
GCLF is symmetric}. Doing so we find that the predictions of the Equations~(\ref{eqn:ll}) to
(\ref{eqn:Tl}) are accurate to within $\sim\!30\%$ for our sample of late-type
dwarf galaxies. 

The above relations can be understood as a minimum GC formation efficiency
threshold. In other words, the inverse scaling relations with galaxy $L_V$
suggest that dwarf galaxies at fainter total luminosities are either more
efficient in forming star clusters relative to the formation of field stars or
less efficient in the formation  of field stars than more luminous systems
(see Fig.~\ref{Fig:SNLMT_compare}). We have discussed above (see Sect.\,\ref{subSectSN}) 
that passive luminosity fading of the host is almost parallel to this lower limit 
and leads to an increase in the $S_N$ value, provided the GC population size 
remains constant. However, it should be noted that the slope observed in the 
data partially reflects the slope stemming from the definition of these equations. 

%%%%%%%%%%%%%%%%%%%%%%%%%%%%%%%%%%%%%%%
\subsection{Globular Cluster System Scaling Relations as a Function of Galaxy Mass}
\label{Sect:GCscaling}

\begin{figure*}
\includegraphics[width=1\textwidth, bb=25 25 710 440]{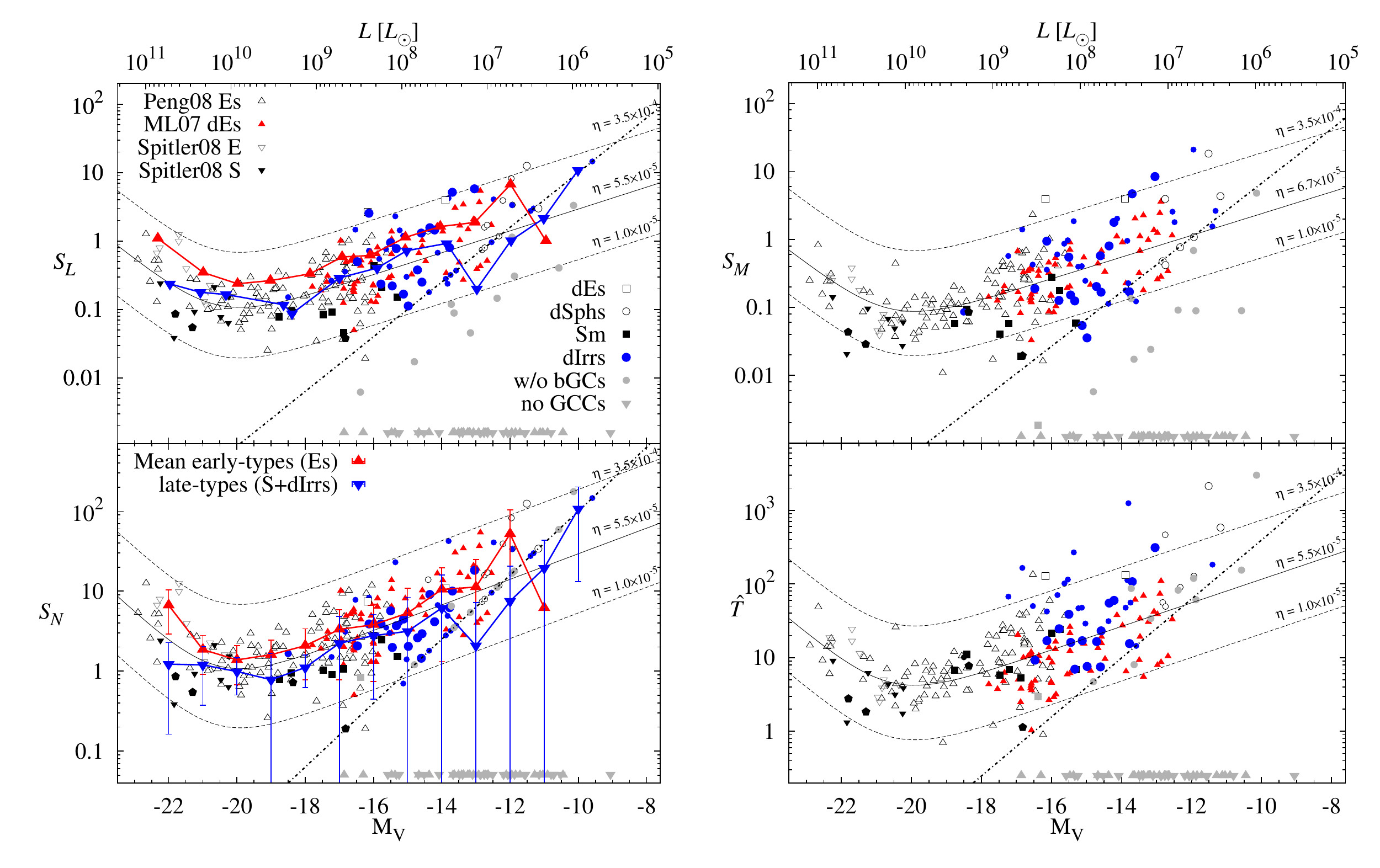}
\caption{Globular cluster system scaling parameters as a function of galaxy 
luminosity. From top left to bottom right are shown the GC specific 
specific luminosities ($S_L$), specific mass ($S_M$), specific frequencies ($S_N$), 
and specific number ($\hat{T}$) for all galaxies in the combined sample. 
Large solid dots indicate dIrrs from our study, while small solid dots are 
dIrrs from \protect\cite{Seth04, Sharina05} and \protect\cite{Georgiev06}. 
Solid pentagons show the corresponding values for the Milky Way, M31, 
LMC and SMC. Grey triangles show galaxies in which no GC candidates are
detected. The dash-dotted line represents the corresponding value if
galaxy hosts one GC at present, see Sect.~\ref{Sect:LowerLimit}. The solid 
curves are predictions by the models which assumes that GCs form proportional 
to the total galaxy halo mass and that stellar, SNe-driven feedback and virial 
shock-heating regulate galaxy wide star formation \protect\citep{Dekel&Birnboim06}, 
below and above ${\cal M}\simeq3\times10^{10}{\cal M}_\odot$, respectively (see 
Sect.\,\ref{Sect:GCscaling}). The solid curve (top panels) is the best-fit $\eta_L=5.5$ 
and $\eta_M=6.7\times10^{-5}$ value to the $S_L$ and $S_M$ distributions, 
respectively. The solid lines in the bottom panels are curves with adopted 
$\eta_L=5.5\times10^{-5}$. Large triangles connected with solid lines  
in the left panels show the co-added running average $S_N$ and $S_L$ values 
as a function of galaxy luminosity for early- and late-type systems. Dashed 
curves illustrate the range in $\eta$ values among galaxies, between 
$\eta = 10^{-5}$ to $3.5\times10^{-4}$.
\label{Fig:SNLMT_compare}}
\end{figure*}

All panels in 
Figure~\ref{Fig:SNLMT_compare} strengthen the previously observed increasing
GC formation efficiency with decreasing galaxy luminosity for galaxies
fainter than $M_V\!\approx\!-20$ mag. This trend is preserved even when
the GC number (or mass) is compared to the {\it total} host galaxy mass.
Theoretical models of star-formation processes regulated by supernova
feedback in low-mass halos \cite[e.g.][]{Dekel&Silk86,Dekel&Woo03},
predict an increasing total ${\cal M}_h/L$ with decreasing halo mass. The
assumption that the number of GCs could be a function of host galaxy
mass-to-light ratio and environmental density have been investigated in
previous studies of early-type galaxies \citep[e.g.][and references
therein]{Forbes05, Bekki06, Peng08}.

In the following we derive analytical expressions which relate the
observed {\it specific formation efficiency\/} of old GCs in
galaxies as a function of {\it total\/} halo mass ${\cal M}_h$. Under the
assumption that GCs form in proportion to the mass of the host galaxy, 
i.e. the total GC system mass ${\cal M}_{\rm GCS}$ is directly related to 
${\cal M}_h$ as:
\begin{equation}\label{eqn:Mgcs=etaMh}
{\cal M}_{\rm GCS} \equiv \eta {\cal M}_h,
\end{equation}
where $\eta$ represents an empirical formation efficiency parameter,
which is the {\it specific globular cluster formation efficiency}.
This parameter describes the total observed mass fraction in globular
clusters, which is a product of the initial cluster formation efficiency 
(initial cluster mass function) and various mechanisms of cluster dissolution 
(e.g. two-body relaxation, GMC encounters, tidal shocking and dynamical friction).
Note that our sample dwarf galaxies have old GCs with ages expected from their 
$V-I$ colour of $t \ga 4$ Gyr, hence, we are probing the early, post-violent 
relaxation phases of GC formation and their dissolution history with respect 
the field stellar population. In the following, we assume that the total GC
system mass and its corresponding total luminosity are related through
\begin{equation}\label{eqn:Ngc=Lgcs}
L_{\rm GCS}={\cal M}_{\rm GCS}/\gamma_V
\end{equation}
where $\gamma_V$ is the mean $V$-band GC mass-to-light ratio. 

The observed GC system scaling relations have been tied by
previous studies \cite[e.g.][]{Forbes05} into the galaxy formation context
using the models of \cite{Dekel&Silk86} according to which star formation
in dark-matter halos below a stellar mass of ${\cal M}\approx
3\times10^{10}M_\odot$ (${\cal M}_h\sim10^{12}M_\odot$) is governed 
by the thermal properties of the inflowing
gas which are primarily regulated by supernova (SN) feedback. These models
predict that the mass-to-light ratio evolution of mini-halos follows ${\cal 
M}_h/L_V\!\propto\!{\cal M}_h^{-2/3}$ \citep{Dekel&Woo03}, so that
$L_V\!\propto\!{\cal M}_h^{5/3}$.

The latest models on star formation regulated by shock-heating \
\citep{Dekel&Birnboim06} predict that above this characteristic stellar mass 
(${\cal M}\!\approx\!3\times10^{10}{\cal M}_\odot$), the gas inflow 
experiences a virial shock (``hot stream'') and, hence, is more susceptible 
to feedback leading to suppressing star formation. This physical transition leads
to a different mass-to-light ratio scaling that follows ${\cal
M}_h/L_V\!\propto\!{\cal M}_h^{1/2}$ or $L_V\propto{\cal M}_h^{1/2}$ for
${\cal M}\gtrsim3\times10^{10}{\cal M}_\odot$.

\subsubsection{Normalization of the Scaling Relations}\label{Sect:GCscalingNorm}
In order to obtain a {\it quantitative\/} description of the specific GC formation
efficiency as a function of galaxy halo mass we must normalize these scaling 
relations. We, therefore, assume the following normalization relations
\begin{eqnarray}
\kappa_1 L_V = {\cal M}_h^{5/3}  & \mbox{ for } {\cal M}\!<\!3\times10^{10}{\cal
  M}_\odot \label{eqn:kappa} \\ 
\kappa_2 L_V = {\cal M}_h^{1/2}  & \mbox{ for } {\cal M}>3\times10^{10}{\cal
  M}_\odot.\label{eqn:kappa2} 
\end{eqnarray}
Such a normalization has not been investigated by previous studies
\cite[e.g.][]{Forbes05}. To empirically
calibrate the model predictions in the low-mass, SN-feedback regulated
star formation regime (${\cal M}\!<\!3\times10^{10}M_\odot$), we use the
results from \cite{Strigari08} who measure dynamical masses within the
inner 0.3\,kpc of many Local Group dwarf galaxies over a wide range of
luminosities ($10^3\!-\!10^7L_\odot$). \citeauthor{Strigari08} find that all
dwarfs have a ``universal'' central mass of ${\cal M}_{\rm 0.3
kpc}=10^7M_\odot$. This central dynamical mass can be tied through
high-resolution cold dark matter (CDM) simulations to the total halo mass via
${\cal M}_h\simeq {\cal M}_{\rm 0.3 kpc}^{1/0.35}\times10^{-11}M_\odot$, based on
the models of \cite{Bullock01}. Below the critical mass of 
${\cal M}\!=\!3\times10^{10}{\cal M}_\odot$ this leads to $\kappa_1\!=\!({\cal
M}_{\rm 0.3 kpc}^{1/0.35}\!\times\!10^{-11})^{5/3}/L_V$. 
For $L_V$, we choose $L_V\!=\!10^6L_\odot$ as the lowest luminosity of a
stellar system that can possibly form one GC. Furthermore, the dynamical mass
measurements of \citeauthor{Strigari08} are less dominated by observational 
errors at this luminosity. For the high-mass, ``virial-shock'' regime we base the
determination of $\kappa_2$ on empirical results for the virial mass to $B$-band
light ratio at $3\times10^{10} {\cal M}_\odot$: ${\cal M}_{\rm vir}/L_B=75$
from \cite{Eke06} based on galaxy group dynamics. From SSP models at solar metallicity 
\citep{BC03} we derive $({\cal M}/L_{\rm B})/({\cal M}/L_{\rm V})=1.65$ 
at 14\,Gyr which leads to ${\cal M}_{\rm vir}/L_V=41$. 
Hence, we obtain $L_V=6.67\times10^8{\cal M}_\odot$,
i.e., $\kappa_2=10^{-4.42}$ for the high-mass regime. Thus, we have
\begin{eqnarray}\label{eqn:Mh=cte}
10^9L_V = {\cal M}_h^{5/3} & \mbox{ for } {\cal M}\!<\!3\times10^{10}{\cal M}_\odot \\
10^{-4.42}L_V = {\cal M}_h^{1/2} & \mbox{ for } {\cal M}>3\times10^{10}{\cal M}_\odot.
\end{eqnarray}
Using Equations\,(\ref{eqn:Mgcs=etaMh}) through (\ref{eqn:kappa}) we can
express the scaling relation defined by Equations~(\ref{eqn:sn}), (\ref{eqn:sm}), (\ref{eqn:sl}), and
(\ref{eqn:T}) as a function of galaxy absolute magnitude for the ``SN
feedback'' regime at stellar mass ${\cal M}_h \!<\! 3\times10^{10}{\cal
  M}_\odot$: 
\begin{equation}
\log S_N \simeq 6+0.24\times M_{V,\odot} + \log\left(\frac{\eta\kappa_1^{0.6}}{m_{\rm TO}}\right) + 0.16 M_{V}
\label{eqn:SNevolMv}
\end{equation}

\begin{equation}
\log S_L \simeq 2-0.16\times M_{V,\odot} + \log\left(\frac{\eta\kappa_1^{0.6}}{\gamma_V}\right) + 0.16M_V
\label{eqn:slevolMv}
\end{equation}

\begin{equation}
\log S_M \simeq 2-0.16\times M_{V,\odot} + \log\left(\frac{\eta\kappa_1^{0.6}}{\Upsilon_V}\right) + 0.16M_V
\label{eqn:smevolMv}
\end{equation}

\begin{equation}
\log \hat{T} \simeq 9-0.16\times M_{V,\odot} + \log\left(\frac{\eta\kappa_1^{0.6}}{m_{\rm TO}\Upsilon_V}\right) + 0.16M_V
\label{eqn:tevolMv}
\end{equation}
where $\Upsilon_V$ is the baryonic $V$-band mass-to-light ratio of the
host galaxy, $m_{\rm TO}$ is the GCLF turn-over GC mass, and $\gamma_V$ is
the corresponding GC mass-to-light ratio. The scaling relations as a function
of galaxy luminosity $L_V$, total halo mass ${\cal M}_{\rm h}$, and
baryonic mass ${\cal M}_b$ are given in Appendix\,\ref{appendix}.

For the "virial shock" regime at masses ${\cal M} > 3\times10^{10}
{\cal M}_\odot$ (${\cal M}_h\gtrsim10^{12}{\cal M}_\odot$) the above 
scaling relations have the following functional forms: 

\begin{equation}
\log S_N \simeq 6+0.8\times M_{V,\odot} + \log\left(\frac{\eta\kappa_2^{2}}{m_{\rm TO}}\right) - 0.4 M_V
\label{eqn:SNevolMv2}
\end{equation}

\begin{equation}
\log S_L \simeq 2+0.4\times M_{V,\odot} + \log\left(\frac{\eta\kappa_2^{2}}{\gamma_V}\right) - 0.4 M_V
\label{eqn:slevolMv2}
\end{equation}

\begin{equation}
\log S_M \simeq 2+0.4\times M_{V,\odot} + \log\left(\frac{\eta\kappa_2^{2}}{\Upsilon_V}\right) - 0.4 M_V
\label{eqn:smevolMv2}
\end{equation}

\begin{equation}
\log \hat{T} \simeq 9+0.4\times M_{V,\odot} + 
\log\left(\frac{\eta\kappa_{2}^{2}}{m_{\rm TO}\Upsilon_V}\right) - 0.4 M_V.
\label{eqn:tevolMv2}
\end{equation}

\subsubsection{Continuous description of the GCS scaling relations}\label{Sect.ContinuousRelations}
In order to simultaneously describe the observed trends in the various GCS 
scaling relations in the stellar/SNe-dominated feedback regime {\it and} the 
regime where star-formation is primarily regulated by virial shock-heating of  
the inflowing gas, we combine the above sets of relations above and below the 
characteristic galaxy stellar mass at ${\cal M}\sim3\times10^{10} {\cal M}_\odot$:
\begin{eqnarray}
S_N = \frac{\eta10^6}{m_{\rm TO}}\left(\kappa_1^{0.6}10^{0.16M_V + 0.24M_{V,\odot}} + \kappa_2^{2}10^{-0.4M_V + 0.8M_{V,\odot}}\right)
\label{eqn:CSNevolMv}
\end{eqnarray}
\begin{eqnarray}
S_L = \frac{\eta10^2}{\gamma_V}\left(\kappa_1^{0.6}10^{0.16(M_V - M_{V,\odot})} + \kappa_2^{2}10^{-0.4(M_V - M_{V,\odot})}\right)
\label{eqn:CSLevolMv}
\end{eqnarray}
\begin{eqnarray}
S_M = \frac{\eta10^2}{\Upsilon_V}\left(\kappa_1^{0.6}10^{0.16(M_V - M_{V,\odot})} + \kappa_2^{2}10^{-0.4(M_V - M_{V,\odot})}\right)
\label{eqn:CSMevolMv}
\end{eqnarray}
\begin{eqnarray}
\hat{T} = \frac{\eta10^9}{m_{\rm TO}\Upsilon_V}\left(\kappa_1^{0.6}10^{0.16(M_V - M_{V,\odot})} + \kappa_2^{2}10^{-0.4(M_V - M_{V,\odot})}\right)
\label{eqn:CTevolMv}
\end{eqnarray}

Solid and dotted curves in Figure~\ref{Fig:SNLMT_compare} illustrate the
relations from Eqs.~(\ref{eqn:CSNevolMv}) to (\ref{eqn:CTevolMv}), for 
various values of $\eta$. It can be seen that these scaling relations 
approximate the observed distributions fairly well and indicate a 
statistically significant spread in the mass and luminosity normalized 
GC formation efficiencies $\eta_L$ and $\eta_M$ (top panels in 
Fig.\,\ref{Fig:SNLMT_compare}). The co-added running average $S_L$ 
and $S_N$ values as a function of galaxy luminosity for early- and 
late-type systems (connected with solid line large triangles in the 
left panels of Fig.\,\ref{Fig:SNLMT_compare}) also show that they 
compare well with the analytical model expectation. 

The remarkably good description of the observations by the analytical 
model advocates that at low galaxy masses, where galaxies have shallower 
potential wells (i.e., lower binding energy), gas is effectively heated 
up by the ionizing radiation of SNe suppressing star-formation. In 
the view of the \citeauthor{Dekel&Birnboim06} models star-formation in 
massive and dense molecular clouds is effectively shielded from heating 
radiation which does not significantly influence cluster formation but 
suppresses more effectively the formation of field stars which leads to 
an increase in the GCS scaling parameters. Thus, at low-galaxy masses 
the initial burst of star formation leads to the formation of the old 
metal-poor GCs. The subsequent cooling of the galaxy ISM and its chemical 
enrichment can lead to the formation of metal-rich and younger GCs 
triggered by instability of internal (tidal density waves and/or collisions 
of stellar winds) or external (e.g. galaxy-galaxy interaction) origin. 
Examples are the metal-rich ([Fe/H$]\lesssim-0.5)$ and young $(>$\,few\,Gyr) 
clusters in the Magellanic Clouds and Local Group (LG) dwarf galaxies. 
This is consistent with the recent age-metallicity relation (AMR) derived 
for those systems \citep{Forbes&Bridges10, Sharina10}. These studies 
show that the formation of star clusters in low-mass systems captures 
a snapshot of the enrichment history of the host galaxy which is a 
function of host mass expressed by the metallicity-luminosity relation 
of dSphs/dIrr galaxies \cite[e.g.][]{Mateo98,Crnojevic10,Lianou10}. 
If some of the red GCs in our sample are indeed young (see discussion 
in Sect.\,\ref{Sect:S_N_for_our_sample}) this might support this scenario.

With increasing galaxy mass, the galaxy potential deepens, the binding 
energy is larger, which leads to a more effective field star-formation 
efficiency at the same initial GC formation efficiency, thus causing 
the decrease in scaling parameters, e.g. $S_L$. Beyond the virial 
shock-heating galaxy mass, dense molecular clouds are better shielded 
than the lower-mass, less compact gas clouds in which most of the 
field star-formation will occur. On the one hand, this will lead to 
a nearly universal initial GC formation efficiency, on the other hand, 
to a more efficiently suppressed field star-formation efficiency. 
This will cause the observed upturn in $S_L$ at high galaxy masses. 
That is, at the transitional stellar mass of ${\cal M}\sim3\times10^{10}{\cal M}_\odot$ 
the thermal properties of the inflowing gas are primarily governed 
by virial shocks which heat the ISM and inhibit the efficient formation 
of field stars at the same initial GC formation efficiency. 
The variance between the observed and expected $S_L$ values for this 
model can be in part explained by the interplay of dissipational and 
dissipationless galaxy merging. Observations support the formation of 
more and metal-rich clusters during gas-rich mergers \cite[e.g.][]
{Whitmore&Schweizer95,Whitmore99,Goudfrooij07}. Semi-analytical models 
of cluster formation are able to reproduce the GC metallicity and age 
distributions assuming formation of GCs during galaxy merging \cite[e.g.]
[]{Muratov&Gnedin10}.

%%%%%%%%%%%%%%%%%%%%%%%%%%%%%%%%%%%%%%%
\subsection{The Distribution of Globular Cluster Formation Efficiencies}
\label{Sect:GCefficiency}

\begin{figure*}
\center
\includegraphics[width=1\textwidth, bb= 25 25 710 260]{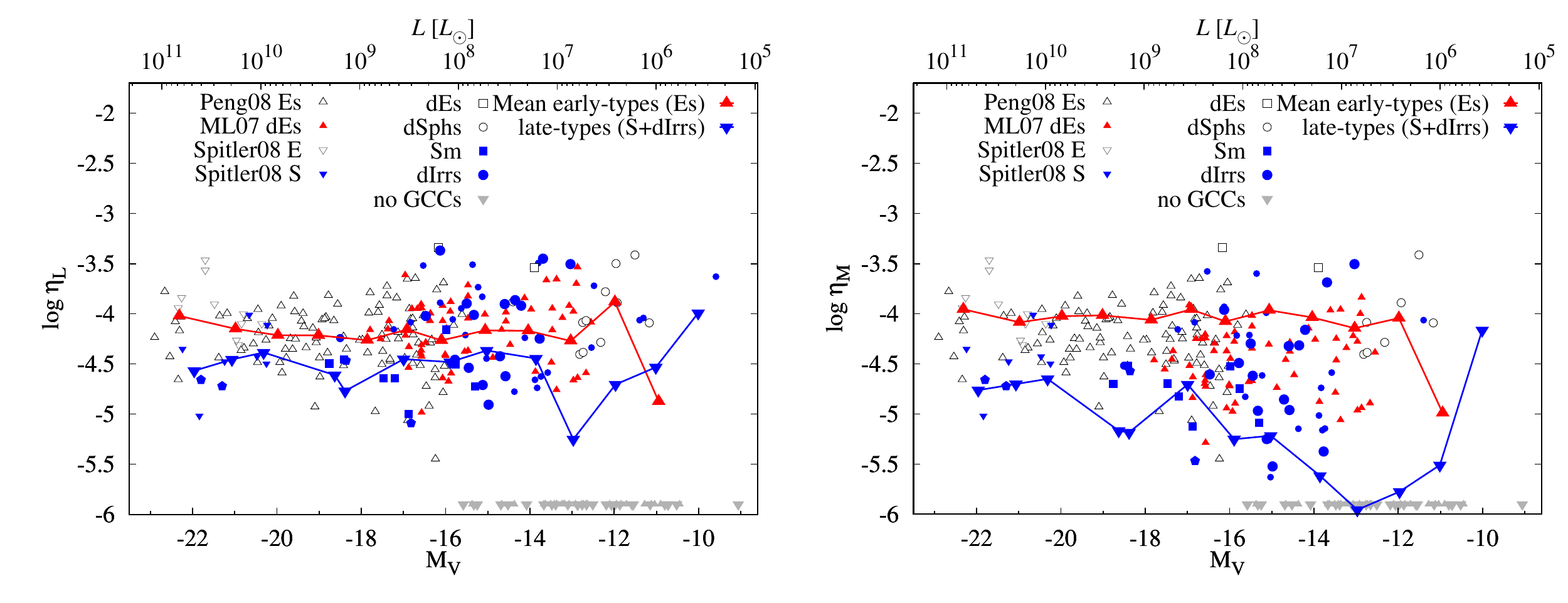}
\caption{Specific globular cluster formation efficiencies, $\eta$, as a 
function of galaxy luminosity. {\bf Left:} Luminosity normalized GC
formation efficiency ($\eta_L$) derived from the $S_L$ relation 
(eqn.\,\ref{eqn:CSLevolMv}). {\bf Right:} Mass normalized GC formation 
efficiency ($\eta_M$) derived from the $S_M$ relation 
(eqn.\,\ref{eqn:CSLevolMv}). The symbols connected with lines show the 
mean $\eta$ values (including galaxies without GCCs) in bins of one 
magnitude. The remaining symbols are as in Figure\,\ref{Fig:SNLMT_compare}.
\label{Fig:Mv-eta}}
\end{figure*}

In the following we investigate the specific GC formation efficiency 
$\eta$ as a function of galaxy luminosity and morphology. 
To derive the distribution of $\eta$ we choose $S_M$ and $S_L$ (upper panels
in Fig.~\ref{Fig:SNLMT_compare}) as their values are least affected by internal 
systematics. Both parameters are independent of distance and are nearly 
insensitive to incompleteness of low-mass, faint GCs
\citep[see Sect.~\ref{SmSect} and also][]{Harris91, Harris01}. This is
reflected in a smaller observed scatter in $S_M$ and $S_L$ values around
the best-fit iso-$\eta$ relation compared to the $S_N$ and $\hat{T}$ values. 
The advantage of the $S_M$ value is that variations in stellar population 
parameters of the host galaxies and GC systems (i.e., their star-formation 
histories) are taken into account by the galaxy $M/L$ (see Sect.~\ref{SmSect}). 
However, we point out that the derived
$\Upsilon_V$ for low-mass dwarf irregular galaxies, which is based on
galaxy colours, is likely to have an increased uncertainty due to their 
likely different SFHs relative to the sample of more massive galaxies from
which the \citeauthor{Bell03} relations were derived (2MASS \& SDSS
galaxies).
We invert Equations~(\ref{eqn:CSLevolMv}) 
and (\ref{eqn:CSMevolMv}) to derive $\eta$ values for our sample galaxies.
The resulting $\eta$ distributions as a function of galaxy luminosity are shown in 
Figure\,\ref{Fig:Mv-eta}. 
To test for trends of the GC formation efficiency as a 
function of galaxy luminosity and morphology we formed the running 
average for $\eta_L$ and $\eta_M$ as a function of galaxy luminosity 
in galaxy luminosity bins of $M_V=1$\,mag, same as in Figure\,\ref{Fig.SNcompare2}. 
The mean values are shown with solid curves and different symbols 
for the two galaxy morphology classes. The errors in the running 
average, not shown for clarity purposes, correspond to approximately 
$^{+1.0}_{-1.5}$\,dex. 

The direct comparison between the $\eta_L$ 
and $\eta_M$ distributions in Figure\,\ref{Fig:Mv-eta} shows good 
agreement for early-type galaxies which indicates a robust understanding 
of their mass-to-light ratios. However, for late-type galaxies the 
systematically lower $\overline{\eta}_M$ with decreasing galaxy 
luminosity is caused by dwarfs without GC candidates. Those GC-free dwarf
galaxies contribute mostly stellar light/mass but no GCs, i.e. the 
normalized GC formation efficiency is reduced at a given galaxy magnitude bin.

Assuming consistent $\overline{\eta}_M$ and $\overline{\eta}_L$ distributions 
for dIrrs this translates into various (degenerate) systematics that are not
included in the models, namely $i)$ a growing mass component towards 
fainter galaxy luminosities, $ii)$ a steeper IMF in low-$L$ dIrr galaxies, or/and $iii)$ 
a systematically higher internal extinction value in low-mass galaxies. 
The investigation of these systematics suggests a mass-dependent term $\alpha({\cal M}_h)$ 
in the scaling relation ${\cal M}_h/L_V\!\propto\!{\cal M}_h^{-2/3 + 
\alpha({\cal M}_h)}$ that is currently unaccounted for in the 
\citeauthor{Dekel&Birnboim06} models. The quantification of such a 
term goes beyond the scope of this paper and requires much more 
comprehensive GCS samples across the entire range of late-type galaxies 
from low-mass dwarf irregulars to massive spirals.

The fits to the $S_L\!-\!M_V$ and $S_M\!-\!M_V$ distributions return 
$\langle\eta_L\rangle\!=\!5.5\times10^{-5}$ and
$\langle\eta_M\rangle\!=\!6.7\times10^{-5}$  
for the luminosity-- and mass-normalized GC formation efficiencies, respectively. 
Those are shown with solid curves in the top panels in
Figure\,\ref{Fig:SNLMT_compare}.  
Because of the uncertainty in the derived galactic $\Upsilon_V$ affecting $\eta_M$ 
we adopted the $\eta_L=5.5\times10^{-5}$ for the bottom panels in 
Figure\,\ref{Fig:SNLMT_compare}. 
The data can be approximated by a wide range of specific GC formation efficiencies from
$\eta\!\simeq\!10^{-5}$ to $\simeq 5\times10^{-4}$. 
The varying scatter in $\eta_L$ as a function of galaxy luminosity is at least
partly due to the increasingly stochastic nature of star and star-cluster
formation processes towards lower galaxy masses. This is supported by the 
observational errors which are lower at higher galaxy luminosities
(cf. Fig.\,\ref{Fig.SN}) which relate to smaller standard deviations of
$\eta_L$ (cf.\ top left panel of Fig.\,\ref{Fig:Mv-eta}). At the faintest 
galaxy luminosities the $\eta_L$ distribution is affected by the
minimum-$\eta_L$ relation (discussed in Sect.~\ref{Sect:LowerLimit}) which
narrows its distribution. The large scatter in $\eta_M$ at faint galaxy
luminosities (relative to that of $\eta_L$) is primarily driven by the poor
knowledge of the galaxy's ${\cal M}/L-$ratios based on optical colours. 

The value of the mean GC formation efficiency, $\eta$, that we obtain 
is consistent with previous estimates by \cite{Blakeslee99, McLaughlin99, 
Kravtsov&Gnedin05, Spitler09}. The value of $\eta_b=1.71\times10^{-4}$ 
obtained by \cite{Blakeslee99} is the GC formation efficiencies with respect 
the baryon galaxy mass (stellar plus hot gas mass) derived for massive 
elliptical galaxies. Multiplying this value by $f_b=0.17$, the universal 
cosmological baryon fraction observed by WMAP \citep{Hinshaw09}, leads to 
$\eta_h=2.9\times10^{-5}$. This value is identical to the $\eta-$values 
obtained by us, \cite{Kravtsov&Gnedin05} and \cite{Spitler09}, 
$\eta_h=2-5\times10^{-5}$ and $7.1\times10^{-5}$, respectively.

\begin{figure}
\center
\includegraphics[width=0.4\textwidth, bb=70 25 450 710]{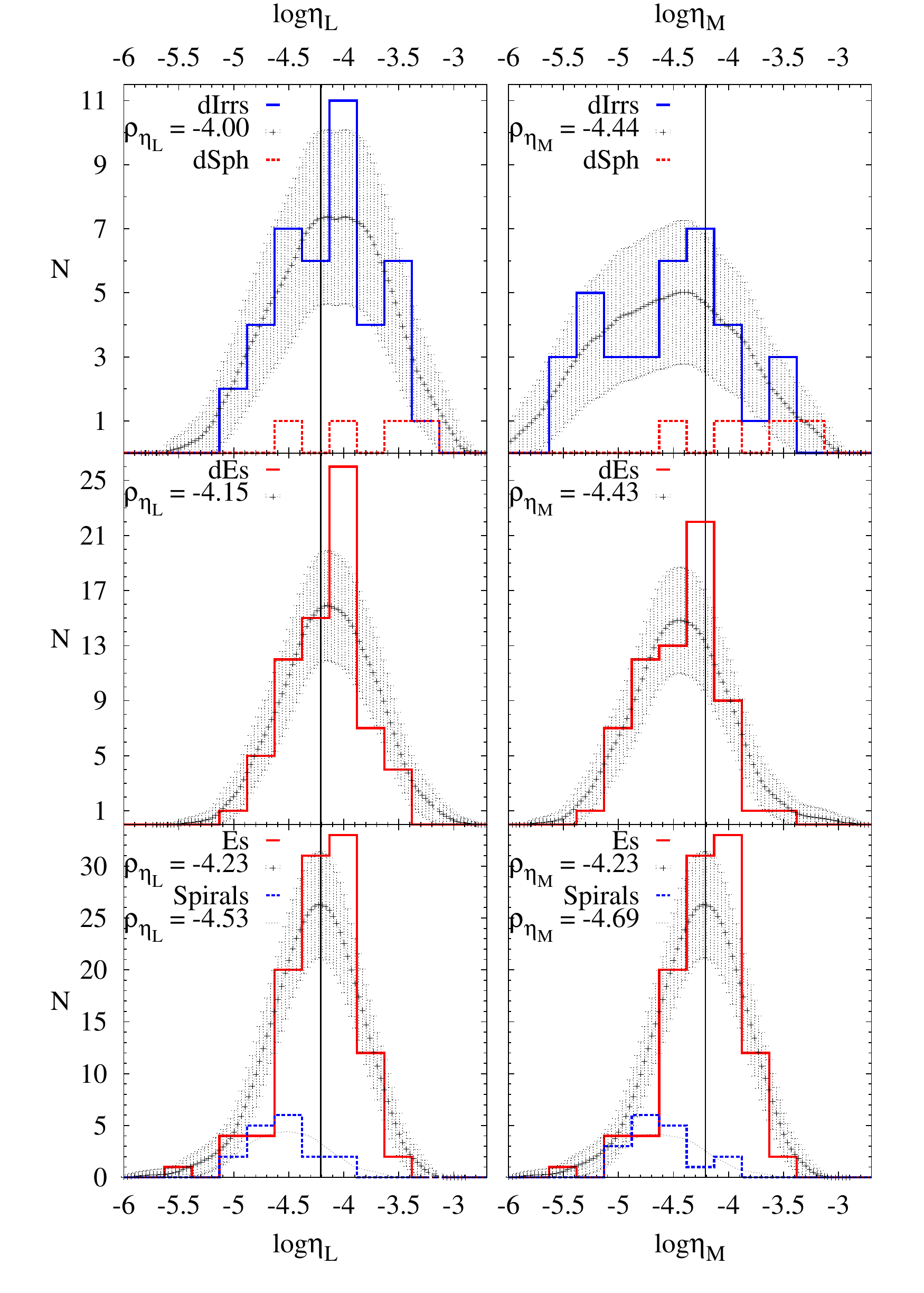}
\caption{Specific GC formation efficiency, $\eta$, for various galaxy types: 
dIrr, dSph, dE, elliptical and spiral galaxies (top to bottom panels). {\it 
Left panels:} luminosity normalized  GC formation efficiency ($\eta_L$) distributions. 
Those for dIrrs are {\it not} corrected for evolutionary fading of the integrated 
galaxy light.  {\it Right panels:} mass normalized GC formation efficiency 
($\eta_M$) distributions. 
Dashed-dotted curves with Poisson errors are Epanechnikov-kernel probability density 
estimates. The maximum-likelihood $\eta$ values for the different galaxy samples 
are shown in the upper left corner of each panel. The vertical solid line in all panels 
shows for comparison the $\eta-$value for Es. Fading of $\Delta M_V=1.5-1$\,mag 
will shift the $\log\eta$ distribution of late-types toward higher values by 
$0.36-0.24$ dex.
\label{Fig:eta-hist}}
\end{figure}

To investigate how $\eta$ relates to different galaxy morphologies we show
histogram distributions in Figure\,\ref{Fig:eta-hist} for the different
morphological types: dIrrs, dSphs, dEs, spirals, and ellipticals. For those 
samples we derive the highest probability density distributions excluding 
the dwarfs with no blue or any GC detection (discussed in Fig.\,\ref{Fig:Mv-eta}). 
The number of such galaxies in the dIrrs, dSphs, dEs, Es and spirals samples, 
respectively, is: 44 (51\%), 4 (60\%), 16 (18\%), 0 and 3 (15\%).
For $\eta_L$, we find no evidence for a statistically significant offset in the
average values $\langle\eta_L\rangle$ between late-type and early-type galaxies 
(see legend in Fig.~\ref{Fig:eta-hist}) even though the scatter around the
average value is larger for the fainter dwarf galaxies as mentioned above. 
We do however find a marginally significant offset between the
$\langle\eta_L\rangle$ values of dIrrs and dEs. The fact that this offset is
larger for $\langle\eta_M\rangle$ than for $\langle\eta_L\rangle$ suggests
that it is mainly due to a systematic difference in stellar population
properties between dIrrs and dEs.  

In general, an offset in $\eta_L$ versus $\eta_M$ between late- and early-type 
galaxies can be expected due to a difference in stellar population parameters
of GCs and  their host galaxies. An increase in $\gamma_V$ by $\sim\!10\%$,
which corresponds  to a few Gyr age difference within the age range of our
sample selection ($t_{\rm GC}\ga 4$ Gyr), introduces a shift in the $\eta_L$
distribution towards larger values ($\Delta\log\eta_L=0.03$), which is
insufficient to account for the observed difference in $\eta$. However, 
luminosity fading of the host galaxy of the order of $\Delta M_V=1.0$ to 1.5\,mag 
(see Sect.~\ref{subSectSN}) would cause $\log\eta_L$ to shift by 0.24 to 
0.36\,dex towards higher values, which would be enough to equalize the 
$\eta_L$ distributions of late-type and early-type galaxies. Although some 
of our sample dIrrs may require such luminosity corrections (in particular 
those with very low $\eta_L$), most of our sample dIrrs lack any evidence 
for significant present-day star formation rates which suggests that evolutionary 
fading corrections of their galaxy light might not be required. On the other 
hand, the mass-normalized GC formation efficiency $\eta_M$ should in principle 
correct for any stellar population age difference between late-and early-type 
galaxies. What we observe instead in Figure\,\ref{Fig:eta-hist} is that the 
late-type galaxies in our sample (dIrrs and spirals) have lower average 
$\eta_M$ than the early-type galaxies at the same luminosity, 
suggesting that the ${\cal M}/L-$ratios of the late-type galaxies may be
underestimated (as already discussed above in the context of 
Fig.\,\ref{Fig:Mv-eta}). However, the number of spiral galaxies in our 
sample is only marginally significant to draw statistically meaningful
conclusions. Therefore, a more comprehensive  study including
near-infrared imaging or spectroscopic observations (lacking for the  
majority of the dwarfs in our sample) should help in resolving this issue. 

Finally, we note that the minimum $\eta$-limit described by
Eq.~(\ref{eqn:ll}) intercepts the iso-$\eta$ line for the maximum
observed specific GC formation efficiency (Eq.~\ref{eqn:CSLevolMv}) at
$M_V\approx-9$ mag, corresponding to $3.37\times10^5L_\odot$ 
(cf.~Fig.~\ref{Fig:SNLMT_compare}). In the context of our assumptions, 
this magnitude describes the lowest luminosity halo that, on average, 
can form and host one old GC. This is consistent with the fact that the 
recently found extremely faint Local Group dwarf galaxies do 
not possess GCs \cite[e.g.][]{Belokurov06, Belokurov07, Irwin07, Walsh07,
Willman05, Willman05b, Zucker06, Zucker06b}.

%%%%%%%%%%%%%%%%%%%%%%%%%%%%%%%%%%%%%%%
\subsection{Implications for Hierarchical Galaxy Formation}
\label{Sect:impl.4_h.g.f.}
\subsubsection{The Influence of various Physical Processes on 
the Variance of Globular Cluster Formation Efficiencies}

The general behavior of the GC system scaling relations implies that 
the star and star-cluster formation processes scale with the total host
galaxy mass. However, it is not yet clear what is driving the large
observed galaxy-to-galaxy scatter in the $\eta$ values for galaxies
with otherwise similar mass. Environmental density is one of the suggested 
parameters to cause systematic trends in this regard \citep{Peng08}. 
The majority of our current dIrr galaxy sample consists of {\it field} 
dIrrs, with a dearth of dIrrs situated in {\it cluster} environments.  
Conversely, dE galaxies with GC data available in our literature
compilation are all located in the Virgo cluster
\cite{Miller&Lotz07,Peng08}.   
Hence we cannot yet make a statistically significant comparison between
the GC formation efficiency of field and cluster dwarf galaxies. To
robustly assess the impact of environmental density on the $\eta$
distribution will require the investigation of more dIrr galaxies in
dense groups and galaxy clusters and more dE galaxies in loose
environments.

Another explanation for the scatter in $\eta$ can be obtained from the 
different timing of star cluster vs.\ field star formation as a function 
of the star formation rate of the host galaxy. For the latter, \cite{Larsen&Richtler00} 
demonstrated that the specific $U$-band luminosity in young massive star 
clusters in late-type galaxies strongly correlates with the total star 
formation rate and star formation rate surface density of the host galaxy, 
SFR and $\Sigma_{\rm SFR}$, respectively. Thus, galaxies with the same 
mass but different SFR histories will have different GC formation efficiencies, 
since galaxies with higher peak SFRs will produce larger numbers of massive 
GCs that will survive a Hubble time.

Furthermore, a delay between the peak formation epochs of star
clusters and field stars that scales with SFR (in the sense that the
star clusters form earlier on average than the field stellar population)
will increase the values of the GC system
scaling parameters, especially for more violent starbursts which likely
form more star {\it clusters\/} per unit gas mass. Any truncation of
star-formation processes before the formation of the field stellar
population is complete (e.g., through gas stripping during infall into
a dense galaxy cluster medium) would further increase the GC scaling 
parameters. A hint that such effects might be at work was provided
by \cite{Peng08} who correlated the SFR and $\Sigma_{\rm SFR}$ with the
mass fraction of GCs in their sample galaxies to find that the cluster
formation rate peaks $\sim\!500$\,Myr earlier than the total galaxy
SFR. 

Another plausible mechanism to account for an apparent variation in 
$\eta$ among galaxies is GC stripping and accretion \citep{Cote98,
Cote02, Hilker99} that can cause a steeper slope of the $S_L-M_V$ 
relation at brighter galaxy luminosities, where the accretion of high-$\eta$ 
dwarf galaxies may be an important mechanism during the early epochs 
of massive galaxy assembly. Such dwarfs may have completed the process 
of star-cluster formation at high redshifts ($z>>2$) and had their field
star-formation processes suppressed or stopped (due to shock heating) 
during their infall through the dense intracluster medium.

\subsubsection{Predictions from a Simplistic Accretion Model}

Some of the most luminous early-type galaxies in our sample show excess 
$\eta$ values ($\log\eta\approx-3.5$) that may indicate that their GCSs 
either experienced an early, highly-efficient episode of massive cluster 
formation followed by a shutoff of (field) star formation or that their 
GCSs were supplemented with high-$\eta$ low-mass dwarf galaxies and their 
GC systems (or a combination of both effects). Such low-mass dwarfs would 
mainly contribute to the blue GC population and rather less to the field 
stellar component, and could well account for the observed high ratio of 
the metal-poor GCs to metal-poor field (halo) stars in massive galaxies 
\citep[e.g.,][]{Harris&Harris02}. \cite{Cote98} and \cite{Hilker99} suggested 
that faint ($M_V\!\approx\!-12$\,mag) dwarfs accreted by cD galaxies can 
indeed provide the right number of GCs to account for their high $S_N$. 
We can make a simple estimate of the number and luminosity of satellite 
galaxies necessary to increase the $S_L$ value of a progenitor galaxy by 
a certain value by simply using the definition of $S_L$ (Eq.\,\ref{eqn:sl}):
\begin{equation}
\label{eqn:SL-Nacc}
S_L = 10^2 \frac{L_{\rm GCS, prg}+N_{\rm acc}\!\times\!L_{\rm GCS, acc}}{L_{V,\rm prg}+N_{\rm acc}\!\times\!L_{V,\rm acc}}.
\end{equation}
The above equation takes into account the increase in luminosity of the 
progenitor galaxy and luminosity of its GCS as a consequence of the accretion.
Assuming that the accretion process occurs in a dissipationless way (i.e. 
no new stars or GCs are formed) and that no GCs are destroyed, the number 
of accreted galaxies can be expressed with the following relation:
\begin{equation}
\label{eqn:Nacc-vs-Mv}
N_{\rm acc}=\frac{\Delta\,S_{L_V}}{S_{L_{V,\rm acc}} - S_{L_{V,\rm prg}} - 
\Delta\,S_{L_V}}\times\frac{L_{V,\rm prg}}{L_{V,\rm acc}},
\end{equation}
\begin{figure}
\includegraphics[width=0.5\textwidth, bb = 20 25 360 260]{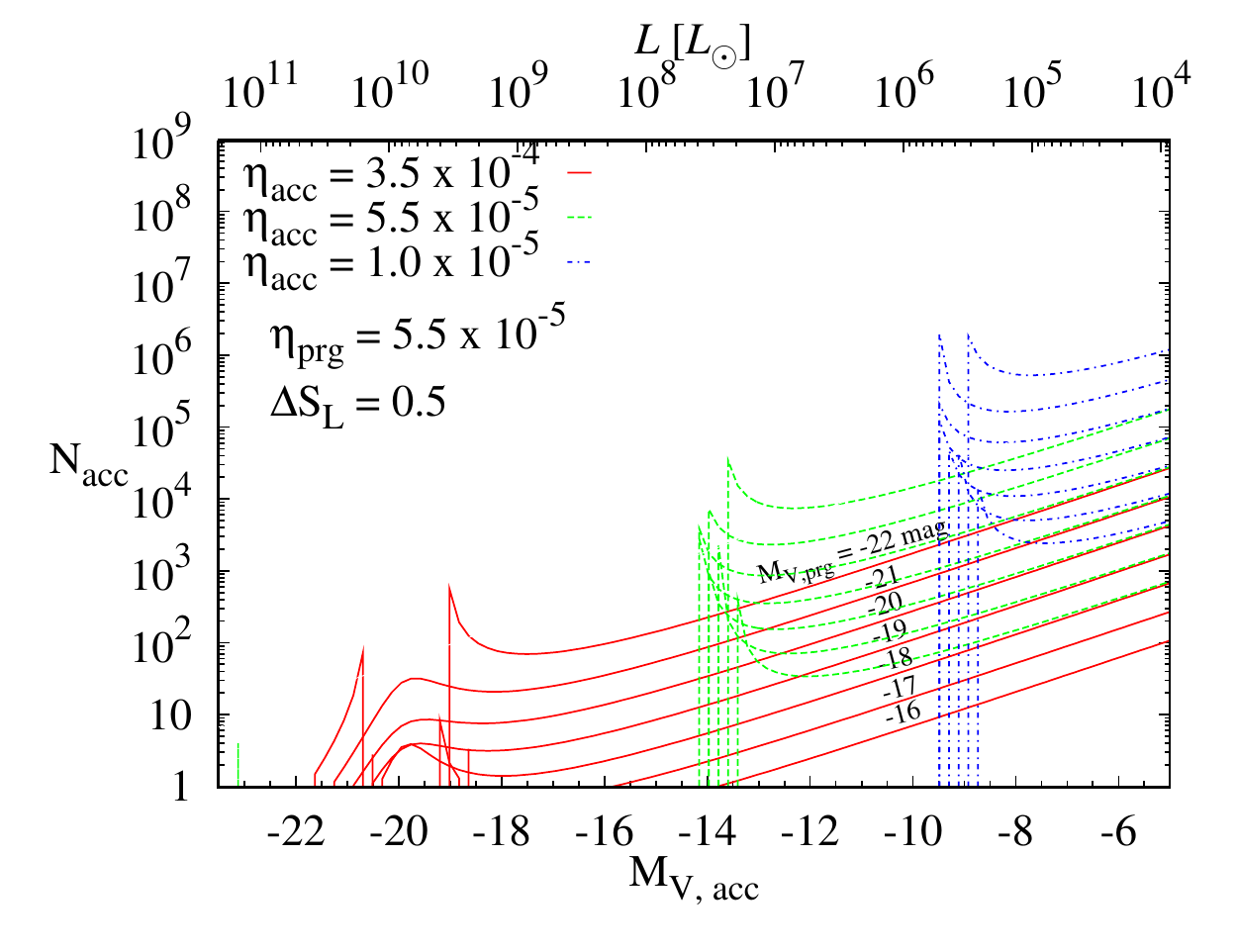}
\caption{An example case for the number of satellite galaxies necessary to boost 
the specific luminosity of a host galaxy by $\Delta S_L=0.5$ as a function 
of satellite luminosity, $M_{V,{\rm acc}}$. The relations are parametrized 
by the luminosity of the progenitor galaxy, $M_{V,{\rm prg}}$. The different 
line-types show three different cases of GC formation efficiencies of the 
accreted satellite galaxy ($\eta_{\rm acc}$). The peaks and "knees" in the 
curves are caused by the inflection point of the "U--shaped" GC system scaling 
relations (cf. Fig.\,\ref{Fig:SNLMT_compare}).
\label{Fig.Nacc-vs-Mv}
}
\end{figure}
where $S_{L_{V,\rm prg}}, L_{V,\rm prg}$ and $S_{L_{V,\rm acc}},
L_{V,\rm acc}$ are the progenitor and accreted galaxy specific luminosity
and brightness, respectively. The increase in the GC system specific 
luminosity due to dissipationless accretion of satellite galaxies is 
$\Delta\,S_{L_V}=S_{L, V}-S_{L_{V,\rm prg}}$. $S_{L, V}$ is the specific 
luminosity of the remnant galaxy after the accretion processes ended. 
Relation \ref{eqn:Nacc-vs-Mv} gives the prescription to compute the 
number of accreted satellite galaxies with a given luminosity and 
$\eta_{\rm acc}$ (cf. top left panel of Figure\,\ref{Fig:SNLMT_compare}) 
necessary to increase $S_L$ by $\Delta S_L$. To illustrate relation 
(\ref{eqn:Nacc-vs-Mv}), we show in Figure\,\ref{Fig.Nacc-vs-Mv} an 
example case for a progenitor galaxy with increased $S_L$ value by 
$\Delta S_L=0.5$ for three different $\eta_{\rm acc}$ values of the 
satellite galaxies and progenitor luminosities in the range 
$M_{V,{\rm prg}}=-16$ to $-22$ mag. The shape and slope of the curves 
is a direct reflection of the functional relation between the $S_L$ 
value and galaxy luminosity observed in Figure\,\ref{Fig:SNLMT_compare} 
and described by eq.\,(\ref{eqn:CSLevolMv}) (see also Eq.\,\ref{eqn:CSLevolLv2}). 
The only purpose of this simple 
example is to show that in the formalism of the current model an 
accretion of low-luminosity dwarfs with high $S_L$ values can increase 
the $S_L$ of a more massive progenitor galaxy. However, only a more 
detailed treatment with sampling an evolving galaxy luminosity function 
will yield realistic number estimates of accreted satellite galaxies 
distributed over a mass spectrum, which is beyond the scope of this study.

Thus, in this simplistic merging picture the accretion of a reasonable 
number of intermediate-luminosity, high-$S_L$ dwarf galaxies offers a 
valid mechanism for boosting the GC system scaling parameters of cD 
galaxies. The observed correlation between the $S_N$ of the cD galaxy 
and the $X-$ray flux of the host cluster \cite{Blakeslee99} supports 
this picture. Accretion and galaxy merging in the cluster core heats 
up the gas leading to the high $X-$ray flux. Such accretion and hierarchical 
growth of cD galaxies is also evidenced by the steepening of the faint-end 
slope of the luminosity function of galaxies in galaxy clusters with 
increasing redshift \cite[e.g.,][]{Ryan07, Khochfar07}. In addition, 
the fraction of low-mass peculiar/irregular galaxies increases with 
redshift \cite[e.g.,][]{Renzini06,Conselice08}. 
Moreover, observations of low-redshift Abell galaxy clusters show that 
galaxy clusters which host cD galaxies have smaller dwarf-to-giant 
ratios than non-cD clusters \citep{Lopez-Cruz97} and a clear radial 
dependence of the faint-end slope of the galaxy luminosity function 
becoming steeper in the cluster outskirts \citep{Barkhouse07}. It 
therefore seems that the contribution of low-mass dwarfs to the buildup 
of the GC system of the most massive galaxies is likely non-negligible, 
especially at high redshifts. In fact, stripping of GCs from present-day 
dwarf galaxies in the cluster environment is suggested by the lack of 
any detectable GC systems around few dEs closest to M\,87 and M\,49 
in the Virgo galaxy cluster \citep{Peng08}. In the context of the 
current cosmological $\Lambda$CDM paradigm of structure formation 
\citep[e.g.,][]{White&Rees78}, it is of great importance to quantify 
the accreted mass fractions of GCs and field stars. It would be highly 
desirable to study the formation and enrichment timescales of the low 
and high-$\eta$ dwarf galaxies in more detail and compare the field 
stellar populations in such galaxies with those of their GC systems. 

%%%%%%%%%%%%%%%%%%%%%%%%%%%%%%%%%%%%%%%
%%%%%%%%%%%%%%%%%%%%%%%%%%%%%%%%%%%%%%%

\section{Conclusions}\label{Sect.Conclusions}

We investigate the much debated behavior of the observed GCS scaling 
parameters as a function of galaxy luminosity, such as the GC specific 
frequency ($S_N$), specific luminosity ($S_L$), specific mass ($S_M$), 
and specific number ($\hat{T}$). Those are the integrated number, 
luminosity mass and specific number of all globular clusters in a galaxy 
normalized to the total galaxy luminosity and/or mass, respectively 
(see Sect.\,\ref{Sect:Analysis}). We derive these quantities from 
{\it HST/ACS} data of low-mass, faint ($M_V\!>\!-16$ mag) dwarf galaxies, 
mainly late-type irregulars, located in the halo regions of nearby 
($\!<\!10$\,Mpc) galaxy groups and in the field \citep{Georgiev08,
Georgiev09}. In order to investigate the scaling relations of their 
GCSs as a function of galaxy luminosity (mass) we also compiled data 
from the literature for massive cluster ellipticals \citep{Peng08} and 
spiral galaxies \citep{Spitler08}. To complement our data in the low-mass 
regime we included 69 cluster dEs  \citep{Miller&Lotz07} and 12 dIrrs 
from \cite{Seth04} and \cite{Georgiev06}, as well as 24 late-type dwarfs 
in the local low-density environment from \cite{Sharina05}. Thus, we 
cover virtually the entire range in galaxy luminosity from $M_V\!=\!-11$ 
to $-23$ mag ($10^6\!-\!10^{11}L_\odot$) where old GCs ($t\!\ga\!4$ Gyr) 
are reliably detected. Our main results can be summarized as follows:\\

\noindent$\bullet$ {\it Trends in GCS scaling relations hold 
irrespective of galaxy morphology.} ---
The significantly increased number of galaxies in our analysis in 
comparison with earlier studies over the entire mass and galaxy 
morphology range allowed us to firmly corroborate the previous 
observational findings that the GCS scaling parameters vary as a 
function of galaxy luminosity (Fig.~\ref{Fig:SNLMT_compare}). We find 
that this relation holds irrespective of galaxy morphological type, 
suggesting a universal mode of GC formation. Galaxies show increasing 
GCS scaling parameters toward low and high-luminosity systems with a 
minimum at around $M_V\!\approx\!-20.5$\,mag ($L_V\approx2\times10^{10}L_\odot$).\\

\noindent$\bullet$ {\it $S_L$ values of early-type galaxies are twice 
that of late-types at a given galaxy luminosity.} ---
The specific luminosity $S_L$ is the most robust scaling parameter and 
shows least scatter due to its independence of distance and weak sensitivity 
to completeness corrections at the faint end slope of the GC luminosity 
function (Sect.\,\ref{Sect:SL}). For late-type galaxies, spirals and 
irregulars the $S_L$ value is on average two times smaller at a given 
galaxy luminosity than that for early-type systems (cf. Fig.\,\ref{Fig.SNcompare2}). 
This difference can be partially accounted for by the passive evolutionary 
fading of the integrated galaxy light ($\Delta V\!\gtrsim\!1$\,mag), 
provided no more GCs are formed or destroyed at the same time (cf. 
arrow in Fig.\,\ref{Fig.SNcompare2}). As cluster dSphs are on average 
fainter and exhibit higher scaling parameters than field dIrrs, the 
above analysis supports the idea that dIrr galaxies could be the 
progenitors of dSphs and fainter dEs \cite[e.g.][]{Miller98}, but see 
also \cite{Grebel03}. However, our sample contains mostly dIrrs  in 
field and group environments. Their GC scaling parameters might differ 
in cluster environments due to varying GC formation efficiencies and/or 
cluster stripping. However, such cluster dIrrs have not been sampled 
well enough yet.\\

\noindent$\bullet$ {\it The "U-shaped" behavior of the GC scaling 
parameters is remarkably well described by a two component model 
at a transitional halo mass critical for the thermal properties of 
the inflowing gas.} ---
In order to explain the trends in the GCS scaling parameters we have
assumed that the total mass in GCs is proportional to the total halo 
mass of the host galaxy (${\cal M}_{\rm GCS}=\eta{\cal M}_{\rm h}$), 
i.e. the formation of GCs scales with the total galaxy mass. The 
coefficient, $\eta$, is the {\it specific GC formation efficiency} 
parameter, which measures the present-day mass/luminosity of the GCS 
drawn from the initial star cluster population mass/luminosity 
function shaped by various mechanisms over a Hubble time (cluster 
relaxation, tidal and dynamical friction, dissolution). We have 
invoked theoretical models of \cite{Dekel&Birnboim06} which predict 
the dependence of the galaxy ${\cal M}/L$ as a function of galaxy 
mass determined by the thermal properties of the inflowing gas for 
two different halo mass regimes in which star formation processes are 
regulated by supernova feedback and virial shocks below and above the 
critical stellar mass of $3\!\times\!10^{10}{\cal M}_\odot$. For 
low-mass halos these models predict that the mass-to-light ratio varies 
with host galaxy mass as ${\cal M}_{h}/L\propto{\cal M}_{h}^{-2/3}$ 
\citep[see also][]{Dekel&Silk86}. To draw quantitative conclusions, 
the absolute normalization of this relation is very important which 
we derive from recent observations of the mass in the inner 0.3\,kpc 
(${\cal M}_{0.3}=10^7{\cal M}_\odot$) of nearby low-mass dwarf galaxies 
\citep{Strigari08} using its relation to the total halo mass from the 
models of \cite{Bullock01}. Above the critical mass, at large halo 
masses the models predict ${\cal M}_{h}/L\propto{\cal M}_{h}^{1/2}$ 
which we calibrate with dynamical mass measurements based on galactic 
group dynamics of massive early-type galaxies \citep{Eke06}. This 
results in an analytical model which describes remarkably well the 
observed trends in the GC scaling relations as a function of galaxy 
luminosity (Sect\,\ref{Sect.ContinuousRelations}). In the \cite{Dekel&Birnboim06} 
model, compact and dense molecular clouds (the sites of GC formation) 
are efficiently shielded against shock heating and ionizing stellar 
feedback above and below the critical galaxy mass, respectively. This 
implies that in this picture, the shape of the GC scaling relations can 
be described as a universal GC formation efficiency for the entire galaxy 
mass range but with an evolving field star-formation efficiency. The 
latter is equally low for low- and high-mass systems (compared to the 
formation of massive/dense star clusters) due to being efficiently 
suppressed by the stellar and SNe ionizing feedback and shock heating 
of the inflowing gas below and above the critical galaxy mass 
($\sim\!10^{10.5}{\cal M}_\odot$), respectively. This causes an evolution 
of the mass in clusters to mass in field stars ratio as a function of 
galaxy mass. The evolution of this ratio is at the roots of the shape 
of the GC scaling relations. The best fit of this model to the observed 
distributions yielded a value of the observed GC formation efficiency 
parameter ($\eta$) for the luminosity and mass-normalized GC formation 
efficiency of $\langle\eta_L\rangle \simeq5.5\times10^{-5}$ and 
$\langle\eta_M\rangle \simeq6.47\times10^{-5}$ (see Sect.\,\ref{Sect:GCefficiency}). 
The $\eta$ distributions are very similar for all galaxy types, if 
passive evolutionary fading of the late-type galaxy sample of $\sim\!1$\,mag 
is applied (cf. Fig.\,\ref{Fig:eta-hist}).\\

\noindent$\bullet$ {\it The differences between model predictions and 
observations can be attributed to a mixture of effects of varying GC 
formation efficiencies, galaxy merging histories, and a variation in 
cluster destruction mechanisms as a function of galaxy mass lacking 
in the current model.} ---
The most massive galaxies, whose $S_M$ and $S_L$ values increase more 
rapidly than expected from the theoretical predictions for a fixed 
$\eta$ can be understood as systems which have either undergone an 
early episode of extremely efficient star-cluster formation or less 
efficient formation of field stars or which have preferentially accreted 
high-$\eta$ dwarf galaxies. In a simple merging picture of satellite 
galaxies of a given luminosity we showed in Sect.\,\ref{Sect:impl.4_h.g.f.} 
that it is possible to boost the $S_L$ values of galaxies by accretion 
of intermediate luminosity, high-$\eta$ dwarfs, thus offering an efficient 
mechanism for explaining the high $\eta$ values of cD galaxies. In 
addition, the observed spread in $\eta$ may be caused by the stochastic 
nature of star and star-cluster formation at low galaxy mass or by 
differences in SFH and SFR intensity for galaxies at the same luminosity.
\\

To understand whether the difference in the GCS scaling-parameter ($S_L$
or $S_M$) distributions between late- and early-type galaxies is
statistically significant and, if so, to understand the nature of this
difference (e.g.~age, chemical enrichment, environment, etc.) it is
crucial to sample with more observations (e.g. near-UV/IR photometry or
spectroscopy) the GCSs of spiral galaxies at intermediate to high masses,
late-type dwarfs in dense cluster environments, both of which are 
significantly under represented in the current study.

%%%%%%%%%%%%%%%%%%%%%%%%%%%%%%%%%%%%%
%%%%%%%%%%%%%%%%%%%%%%%%%%%%%%%%%%%%%

\section*{Acknowledgments}
IG is thankful for the financial support through German Research Foundation 
(\emph{Deut\-sche For\-schungs\-ge\-mein\-schaft, DFG\/}) grant DFG-Projekt 
BO-779/32-1 and by the STScI Director's Discretionary Research Fund. THP 
gratefully acknowledges support from the National Research Council of Canada 
in form of the Plaskett Research Fellowship. The authors would like to thank 
Tom Richtler for providing comments to an earlier version of this paper as 
well as Eric Peng and Knut Olsen for valuable comments and suggestions. The 
authors thank the referee for the constructive input which improved the paper.

\bibliographystyle{aa}
\bibliography{references}

\begin{thebibliography}{128}
\expandafter\ifx\csname natexlab\endcsname\relax\def\natexlab#1{#1}\fi

\bibitem[{{Anders} \& {Fritze-v.~Alvensleben}(2003)}]{Anders03}
{Anders}, P. \& {Fritze-v.~Alvensleben}, U. 2003, A\&A, 401, 1063

\bibitem[{{Ashman} \& {Zepf}(1992)}]{Ashman&Zepf92}
{Ashman}, K.~M. \& {Zepf}, S.~E. 1992, ApJ, 384, 50

\bibitem[{{Ashman} \& {Zepf}(1998)}]{Ashman&Zepf98}
{Ashman}, K.~M. \& {Zepf}, S.~E. 1998, {Globular Cluster Systems} (Cambridge
  University Press, 1998.~(Cambridge astrophysics series ; 30) QB853.5 .A84
  1998)

\bibitem[{{Barkhouse} {et~al.}(2007){Barkhouse}, {Yee}, \&
  {L{\'o}pez-Cruz}}]{Barkhouse07}
{Barkhouse}, W.~A., {Yee}, H.~K.~C., \& {L{\'o}pez-Cruz}, O. 2007, ApJ, 671,
  1471

\bibitem[{{Bassino} {et~al.}(2006{\natexlab{a}}){Bassino}, {Faifer}, {Forte},
  {Dirsch}, {Richtler}, {Geisler}, \& {Schuberth}}]{Bassino06b}
{Bassino}, L.~P., {Faifer}, F.~R., {Forte}, J.~C., {et~al.} 2006{\natexlab{a}},
  A\&A, 451, 789

\bibitem[{{Bassino} {et~al.}(2006{\natexlab{b}}){Bassino}, {Richtler}, \&
  {Dirsch}}]{Bassino06a}
{Bassino}, L.~P., {Richtler}, T., \& {Dirsch}, B. 2006{\natexlab{b}}, MNRAS,
  367, 156

\bibitem[{{Bassino} {et~al.}(2008){Bassino}, {Richtler}, \&
  {Dirsch}}]{Bassino08}
{Bassino}, L.~P., {Richtler}, T., \& {Dirsch}, B. 2008, MNRAS, 386, 1145

\bibitem[{{Bastian}(2008)}]{Bastian08}
{Bastian}, N. 2008, MNRAS, 390, 759

\bibitem[{{Begum} {et~al.}(2008){Begum}, {Chengalur}, {Karachentsev},
  {Sharina}, \& {Kaisin}}]{Begum08}
{Begum}, A., {Chengalur}, J.~N., {Karachentsev}, I.~D., {Sharina}, M.~E., \&
  {Kaisin}, S.~S. 2008, MNRAS, 386, 1667

\bibitem[{{Bekki} {et~al.}(2006){Bekki}, {Yahagi}, \& {Forbes}}]{Bekki06}
{Bekki}, K., {Yahagi}, H., \& {Forbes}, D.~A. 2006, ApJL, 645, L29

\bibitem[{{Bell} {et~al.}(2003){Bell}, {McIntosh}, {Katz}, \&
  {Weinberg}}]{Bell03}
{Bell}, E.~F., {McIntosh}, D.~H., {Katz}, N., \& {Weinberg}, M.~D. 2003, ApJS,
  149, 289

\bibitem[{{Belokurov} {et~al.}(2007){Belokurov}, {Zucker}, {Evans}, {Kleyna},
  {Koposov}, {Hodgkin}, {Irwin}, {Gilmore}, {Wilkinson}, {Fellhauer},
  {Bramich}, {Hewett}, {Vidrih}, {De Jong}, {Smith}, {Rix}, {Bell}, {Wyse},
  {Newberg}, {Mayeur}, {Yanny}, {Rockosi}, {Gnedin}, {Schneider}, {Beers},
  {Barentine}, {Brewington}, {Brinkmann}, {Harvanek}, {Kleinman}, {Krzesinski},
  {Long}, {Nitta}, \& {Snedden}}]{Belokurov07}
{Belokurov}, V., {Zucker}, D.~B., {Evans}, N.~W., {et~al.} 2007, ApJ, 654, 897

\bibitem[{{Belokurov} {et~al.}(2006){Belokurov}, {Zucker}, {Evans},
  {Wilkinson}, {Irwin}, {Hodgkin}, {Bramich}, {Irwin}, {Gilmore}, {Willman},
  {Vidrih}, {Newberg}, {Wyse}, {Fellhauer}, {Hewett}, {Cole}, {Bell}, {Beers},
  {Rockosi}, {Yanny}, {Grebel}, {Schneider}, {Lupton}, {Barentine},
  {Brewington}, {Brinkmann}, {Harvanek}, {Kleinman}, {Krzesinski}, {Long},
  {Nitta}, {Smith}, \& {Snedden}}]{Belokurov06}
{Belokurov}, V., {Zucker}, D.~B., {Evans}, N.~W., {et~al.} 2006, ApJL, 647,
  L111

\bibitem[{{Blakeslee}(1999)}]{Blakeslee99}
{Blakeslee}, J.~P. 1999, AJ, 118, 1506

\bibitem[{{Blakeslee} {et~al.}(1997){Blakeslee}, {Tonry}, \&
  {Metzger}}]{Blakeslee97}
{Blakeslee}, J.~P., {Tonry}, J.~L., \& {Metzger}, M.~R. 1997, AJ, 114, 482

\bibitem[{{Brodie} \& {Strader}(2006)}]{Brodie06}
{Brodie}, J.~P. \& {Strader}, J. 2006, ARA\&A, 44, 193

\bibitem[{{Bruzual} \& {Charlot}(2003)}]{BC03}
{Bruzual}, G. \& {Charlot}, S. 2003, MNRAS, 344, 1000

\bibitem[{{Bullock} {et~al.}(2001){Bullock}, {Kolatt}, {Sigad}, {Somerville},
  {Kravtsov}, {Klypin}, {Primack}, \& {Dekel}}]{Bullock01}
{Bullock}, J.~S., {Kolatt}, T.~S., {Sigad}, Y., {et~al.} 2001, MNRAS, 321, 559

\bibitem[{{Chandar} {et~al.}(2004){Chandar}, {Whitmore}, \& {Lee}}]{Chandar04}
{Chandar}, R., {Whitmore}, B., \& {Lee}, M.~G. 2004, ApJ, 611, 220

\bibitem[{{Conselice} {et~al.}(2008){Conselice}, {Rajgor}, \&
  {Myers}}]{Conselice08}
{Conselice}, C.~J., {Rajgor}, S., \& {Myers}, R. 2008, MNRAS, 446

\bibitem[{{C{\^o}t{\'e}} {et~al.}(1998){C{\^o}t{\'e}}, {Marzke}, \&
  {West}}]{Cote98}
{C{\^o}t{\'e}}, P., {Marzke}, R.~O., \& {West}, M.~J. 1998, ApJ, 501, 554

\bibitem[{{C{\^o}t{\'e}} {et~al.}(2002){C{\^o}t{\'e}}, {West}, \&
  {Marzke}}]{Cote02}
{C{\^o}t{\'e}}, P., {West}, M.~J., \& {Marzke}, R.~O. 2002, ApJ, 567, 853

\bibitem[{{Cox}(2000)}]{Cox00}
{Cox}, A.~N. 2000, {Allen's astrophysical quantities} (Allen's Astrophysical
  Quantities)

\bibitem[{{Crnojevi{\'c}} {et~al.}(2010){Crnojevi{\'c}}, {Grebel}, \&
  {Koch}}]{Crnojevic10}
{Crnojevi{\'c}}, D., {Grebel}, E.~K., \& {Koch}, A. 2010, ArXiv:1002.0341

\bibitem[{{Dahlem} {et~al.}(2006){Dahlem}, {Lisenfeld}, \& {Rossa}}]{Dahlem06}
{Dahlem}, M., {Lisenfeld}, U., \& {Rossa}, J. 2006, A\&A, 457, 121

\bibitem[{{Dekel} \& {Birnboim}(2006)}]{Dekel&Birnboim06}
{Dekel}, A. \& {Birnboim}, Y. 2006, MNRAS, 368, 2

\bibitem[{{Dekel} \& {Silk}(1986)}]{Dekel&Silk86}
{Dekel}, A. \& {Silk}, J. 1986, ApJ, 303, 39

\bibitem[{{Dekel} \& {Woo}(2003)}]{Dekel&Woo03}
{Dekel}, A. \& {Woo}, J. 2003, MNRAS, 344, 1131

\bibitem[{{Dirsch} {et~al.}(2003){Dirsch}, {Richtler}, {Geisler}, {Forte},
  {Bassino}, \& {Gieren}}]{Dirsch03}
{Dirsch}, B., {Richtler}, T., {Geisler}, D., {et~al.} 2003, AJ, 125, 1908

\bibitem[{{Dirsch} {et~al.}(2005){Dirsch}, {Schuberth}, \&
  {Richtler}}]{Dirsch05}
{Dirsch}, B., {Schuberth}, Y., \& {Richtler}, T. 2005, A\&A, 433, 43

\bibitem[{{Durrell} {et~al.}(1996){Durrell}, {Harris}, {Geisler}, \&
  {Pudritz}}]{Durrell96}
{Durrell}, P.~R., {Harris}, W.~E., {Geisler}, D., \& {Pudritz}, R.~E. 1996, AJ,
  112, 972

\bibitem[{{Eke} {et~al.}(2006){Eke}, {Baugh}, {Cole}, {Frenk}, \&
  {Navarro}}]{Eke06}
{Eke}, V.~R., {Baugh}, C.~M., {Cole}, S., {Frenk}, C.~S., \& {Navarro}, J.~F.
  2006, MNRAS, 370, 1147

\bibitem[{{Elmegreen}(2010)}]{Elmegreen10}
{Elmegreen}, B.~G. 2010, ApJL, 712, L184

\bibitem[{{Fall} \& {Zhang}(2001)}]{Fall&Zhang01}
{Fall}, S.~M. \& {Zhang}, Q. 2001, ApJ, 561, 751

\bibitem[{{Forbes}(2005)}]{Forbes05}
{Forbes}, D.~A. 2005, ApJL, 635, L137

\bibitem[{{Forbes} \& {Bridges}(2010)}]{Forbes&Bridges10}
{Forbes}, D.~A. \& {Bridges}, T. 2010, ArXiv:1001.4289

\bibitem[{{Forbes} {et~al.}(1997){Forbes}, {Brodie}, \& {Grillmair}}]{Forbes97}
{Forbes}, D.~A., {Brodie}, J.~P., \& {Grillmair}, C.~J. 1997, AJ, 113, 1652

\bibitem[{{Forbes} {et~al.}(2001){Forbes}, {Georgakakis}, \&
  {Brodie}}]{Forbes01}
{Forbes}, D.~A., {Georgakakis}, A.~E., \& {Brodie}, J.~P. 2001, MNRAS, 325,
  1431

\bibitem[{{Forte} {et~al.}(2001){Forte}, {Geisler}, {Ostrov}, {Piatti}, \&
  {Gieren}}]{Forte01}
{Forte}, J.~C., {Geisler}, D., {Ostrov}, P.~G., {Piatti}, A.~E., \& {Gieren},
  W. 2001, AJ, 121, 1992

\bibitem[{{Fukazawa} {et~al.}(2006){Fukazawa}, {Botoya-Nonesa}, {Pu}, {Ohto},
  \& {Kawano}}]{Fukazawa06}
{Fukazawa}, Y., {Botoya-Nonesa}, J.~G., {Pu}, J., {Ohto}, A., \& {Kawano}, N.
  2006, ApJ, 636, 698

\bibitem[{{Georgiev} {et~al.}(2008){Georgiev}, {Goudfrooij}, {Puzia}, \&
  {Hilker}}]{Georgiev08}
{Georgiev}, I.~Y., {Goudfrooij}, P., {Puzia}, T.~H., \& {Hilker}, M. 2008, AJ,
  135, 1858

\bibitem[{{Georgiev} {et~al.}(2006){Georgiev}, {Hilker}, {Puzia},
  {Chanam{\'e}}, {Mieske}, {Goudfrooij}, {Reisenegger}, \&
  {Infante}}]{Georgiev06}
{Georgiev}, I.~Y., {Hilker}, M., {Puzia}, T.~H., {et~al.} 2006, A\&A, 452, 141

\bibitem[{{Georgiev} {et~al.}(2009{\natexlab{a}}){Georgiev}, {Puzia}, {Hilker},
  \& {Goudfrooij}}]{Georgiev09}
{Georgiev}, I.~Y., {Puzia}, T.~H., {Hilker}, M., \& {Goudfrooij}, P.
  2009{\natexlab{a}}, MNRAS, 392, 879

\bibitem[{{Georgiev} {et~al.}(2009{\natexlab{b}}){Georgiev}, {sHilker},
  {Puzia}, {Goudfrooij}, \& {Baumgardt}}]{Georgiev09b}
{Georgiev}, I.~Y., {sHilker}, M., {Puzia}, T.~H., {Goudfrooij}, P., \&
  {Baumgardt}, H. 2009{\natexlab{b}}, MNRAS, 396, 1075

\bibitem[{{Giavalisco} {et~al.}(2004){Giavalisco}, {Ferguson}, {Koekemoer},
  {Dickinson}, {Alexander}, {Bauer}, {Bergeron}, {Biagetti}, {Brandt},
  {Casertano}, {Cesarsky}, {Chatzichristou}, {Conselice}, {Cristiani}, {Da
  Costa}, {Dahlen}, {de Mello}, {Eisenhardt}, {Erben}, {Fall}, {Fassnacht},
  {Fosbury}, {Fruchter}, {Gardner}, {Grogin}, {Hook}, {Hornschemeier}, {Idzi},
  {Jogee}, {Kretchmer}, {Laidler}, {Lee}, {Livio}, {Lucas}, {Madau},
  {Mobasher}, {Moustakas}, {Nonino}, {Padovani}, {Papovich}, {Park},
  {Ravindranath}, {Renzini}, {Richardson}, {Riess}, {Rosati}, {Schirmer},
  {Schreier}, {Somerville}, {Spinrad}, {Stern}, {Stiavelli}, {Strolger},
  {Urry}, {Vandame}, {Williams}, \& {Wolf}}]{Gavalisco04}
{Giavalisco}, M., {Ferguson}, H.~C., {Koekemoer}, A.~M., {et~al.} 2004, ApJL,
  600, L93

\bibitem[{{Gieles} \& {Bastian}(2008)}]{Gieles&Bastian08}
{Gieles}, M. \& {Bastian}, N. 2008, A\&A, 482, 165

\bibitem[{{Gieles} {et~al.}(2006){Gieles}, {Portegies Zwart}, {Baumgardt},
  {Athanassoula}, {Lamers}, {Sipior}, \& {Leenaarts}}]{Gieles06}
{Gieles}, M., {Portegies Zwart}, S.~F., {Baumgardt}, H., {et~al.} 2006, MNRAS,
  371, 793

\bibitem[{{G{\'o}mez} \& {Richtler}(2004)}]{Gomez&Richtler04}
{G{\'o}mez}, M. \& {Richtler}, T. 2004, A\&A, 415, 499

\bibitem[{{Goudfrooij} {et~al.}(2004){Goudfrooij}, {Gilmore}, {Whitmore}, \&
  {Schweizer}}]{Goudfrooij04}
{Goudfrooij}, P., {Gilmore}, D., {Whitmore}, B.~C., \& {Schweizer}, F. 2004,
  ApJ, 613, L121

\bibitem[{{Goudfrooij} {et~al.}(2007){Goudfrooij}, {Schweizer}, {Gilmore}, \&
  {Whitmore}}]{Goudfrooij07}
{Goudfrooij}, P., {Schweizer}, F., {Gilmore}, D., \& {Whitmore}, B.~C. 2007,
  AJ, 133, 2737

\bibitem[{{Goudfrooij} {et~al.}(2003){Goudfrooij}, {Strader}, {Brenneman},
  {Kissler-Patig}, {Minniti}, \& {Edwin Huizinga}}]{Goudfrooij03}
{Goudfrooij}, P., {Strader}, J., {Brenneman}, L., {et~al.} 2003, MNRAS, 343,
  665

\bibitem[{{Grebel} {et~al.}(2003){Grebel}, {Gallagher}, \&
  {Harbeck}}]{Grebel03}
{Grebel}, E.~K., {Gallagher}, III, J.~S., \& {Harbeck}, D. 2003, AJ, 125, 1926

\bibitem[{{Harris}(1991)}]{Harris91}
{Harris}, W.~E. 1991, ARA\&A, 29, 543

\bibitem[{{Harris}(2001)}]{Harris01}
{Harris}, W.~E. 2001, in Saas-Fee Advanced Course 28: Star Clusters, ed.
  L.~{Labhardt} \& B.~{Binggeli}, 223--+

\bibitem[{{Harris}(2003)}]{Harris03}
{Harris}, W.~E. 2003, in Extragalactic Globular Cluster Systems, ed.
  M.~{Kissler-Patig}, 317--+

\bibitem[{{Harris} \& {Harris}(2002)}]{Harris&Harris02}
{Harris}, W.~E. \& {Harris}, G.~L.~H. 2002, AJ, 123, 3108

\bibitem[{{Harris} {et~al.}(2009){Harris}, {Kavelaars}, {Hanes}, {Pritchet}, \&
  {Baum}}]{Harris09}
{Harris}, W.~E., {Kavelaars}, J.~J., {Hanes}, D.~A., {Pritchet}, C.~J., \&
  {Baum}, W.~A. 2009, AJ, 137, 3314

\bibitem[{{Harris} \& {van den Bergh}(1981)}]{Harris&vdBergh81}
{Harris}, W.~E. \& {van den Bergh}, S. 1981, AJ, 86, 1627

\bibitem[{{Harris} {et~al.}(2006){Harris}, {Whitmore}, {Karakla}, {Oko{\'n}},
  {Baum}, {Hanes}, \& {Kavelaars}}]{Harris06a}
{Harris}, W.~E., {Whitmore}, B.~C., {Karakla}, D., {et~al.} 2006, ApJ, 636, 90

\bibitem[{{Hilker} {et~al.}(1999){Hilker}, {Infante}, \& {Richtler}}]{Hilker99}
{Hilker}, M., {Infante}, L., \& {Richtler}, T. 1999, A\&AS, 138, 55

\bibitem[{{Hinshaw} {et~al.}(2009){Hinshaw}, {Weiland}, {Hill}, {Odegard},
  {Larson}, {Bennett}, {Dunkley}, {Gold}, {Greason}, {Jarosik}, {Komatsu},
  {Nolta}, {Page}, {Spergel}, {Wollack}, {Halpern}, {Kogut}, {Limon}, {Meyer},
  {Tucker}, \& {Wright}}]{Hinshaw09}
{Hinshaw}, G., {Weiland}, J.~L., {Hill}, R.~S., {et~al.} 2009, ApJS, 180, 225

\bibitem[{{Humphrey} {et~al.}(2006){Humphrey}, {Buote}, {Gastaldello},
  {Zappacosta}, {Bullock}, {Brighenti}, \& {Mathews}}]{Humphrey06}
{Humphrey}, P.~J., {Buote}, D.~A., {Gastaldello}, F., {et~al.} 2006, ApJ, 646,
  899

\bibitem[{{Irwin} {et~al.}(2007){Irwin}, {Belokurov}, {Evans}, {Ryan-Weber},
  {de Jong}, {Koposov}, {Zucker}, {Hodgkin}, {Gilmore}, {Prema}, {Hebb},
  {Begum}, {Fellhauer}, {Hewett}, {Kennicutt}, {Wilkinson}, {Bramich},
  {Vidrih}, {Rix}, {Beers}, {Barentine}, {Brewington}, {Harvanek},
  {Krzesinski}, {Long}, {Nitta}, \& {Snedden}}]{Irwin07}
{Irwin}, M.~J., {Belokurov}, V., {Evans}, N.~W., {et~al.} 2007, ApJL, 656, L13

\bibitem[{{Jord{\'a}n} {et~al.}(2007){Jord{\'a}n}, {McLaughlin},
  {C{\^o}t{\'e}}, {Ferrarese}, {Peng}, {Mei}, {Villegas}, {Merritt}, {Tonry},
  \& {West}}]{Jordan07}
{Jord{\'a}n}, A., {McLaughlin}, D.~E., {C{\^o}t{\'e}}, P., {et~al.} 2007, ApJS,
  171, 101

\bibitem[{{Karachentsev} {et~al.}(2006){Karachentsev}, {Dolphin}, {Tully},
  {Sharina}, {Makarova}, {Makarov}, {Karachentseva}, {Sakai}, \&
  {Shaya}}]{Karachentsev06}
{Karachentsev}, I.~D., {Dolphin}, A., {Tully}, R.~B., {et~al.} 2006, AJ, 131,
  1361

\bibitem[{{Karachentsev} {et~al.}(2007){Karachentsev}, {Tully}, {Dolphin},
  {Sharina}, {Makarova}, {Makarov}, {Sakai}, {Shaya}, {Kashibadze},
  {Karachentseva}, \& {Rizzi}}]{Karachentsev07}
{Karachentsev}, I.~D., {Tully}, R.~B., {Dolphin}, A., {et~al.} 2007, AJ, 133,
  504

\bibitem[{{Khochfar} {et~al.}(2007){Khochfar}, {Silk}, {Windhorst}, \&
  {Ryan}}]{Khochfar07}
{Khochfar}, S., {Silk}, J., {Windhorst}, R.~A., \& {Ryan}, Jr., R.~E. 2007,
  ApJL, 668, L115

\bibitem[{{Kissler-Patig}(2000)}]{Kissler-Patig00}
{Kissler-Patig}, M. 2000, in Reviews in Modern Astronomy, ed. R.~E.
  {Schielicke}, 13--+

\bibitem[{{Kravtsov} \& {Gnedin}(2005)}]{Kravtsov&Gnedin05}
{Kravtsov}, A.~V. \& {Gnedin}, O.~Y. 2005, ApJ, 623, 650

\bibitem[{{Kruijssen} \& {Portegies Zwart}(2009)}]{Kruijssen&Zwart09}
{Kruijssen}, J.~M.~D. \& {Portegies Zwart}, S.~F. 2009, ApJL, 698, L158

\bibitem[{{Lada} \& {Lada}(2003)}]{Lada&Lada03}
{Lada}, C.~J. \& {Lada}, E.~A. 2003, ARA\&A, 41, 57

\bibitem[{{Lamers} \& {Gieles}(2008)}]{Lamers&Gieles08}
{Lamers}, H.~J.~G.~L.~M. \& {Gieles}, M. 2008, in Astronomical Society of the
  Pacific Conference Series, Vol. 388, Mass Loss from Stars and the Evolution
  of Stellar Clusters, ed. A.~{de Koter}, L.~J. {Smith}, \& L.~B.~F.~M.
  {Waters}, 367--+

\bibitem[{{Larsen} \& {Richtler}(2000)}]{Larsen&Richtler00}
{Larsen}, S.~S. \& {Richtler}, T. 2000, A\&A, 354, 836

\bibitem[{{Leitherer} {et~al.}(1999){Leitherer}, {Schaerer}, {Goldader},
  {Gonz{\'a}lez Delgado}, {Robert}, {Kune}, {de Mello}, {Devost}, \&
  {Heckman}}]{Leitherer99}
{Leitherer}, C., {Schaerer}, D., {Goldader}, J.~D., {et~al.} 1999, ApJS, 123, 3

\bibitem[{{Lianou} {et~al.}(2010){Lianou}, {Grebel}, \& {Koch}}]{Lianou10}
{Lianou}, S., {Grebel}, E.~K., \& {Koch}, A. 2010, ArXiv:1003.0861

\bibitem[{{Lopez-Cruz} {et~al.}(1997){Lopez-Cruz}, {Yee}, {Brown}, {Jones}, \&
  {Forman}}]{Lopez-Cruz97}
{Lopez-Cruz}, O., {Yee}, H.~K.~C., {Brown}, J.~P., {Jones}, C., \& {Forman}, W.
  1997, ApJL, 475, L97+

\bibitem[{{Maraston}(2005)}]{Maraston05}
{Maraston}, C. 2005, MNRAS, 362, 799

\bibitem[{{Mateo}(1998)}]{Mateo98}
{Mateo}, M.~L. 1998, ARA\&A, 36, 435

\bibitem[{{McLaughlin}(1999)}]{McLaughlin99}
{McLaughlin}, D.~E. 1999, AJ, 117, 2398

\bibitem[{{McLaughlin} \& {Fall}(2008)}]{McLaughlin&Fall08}
{McLaughlin}, D.~E. \& {Fall}, S.~M. 2008, ApJ, 679, 1272

\bibitem[{{McLaughlin} \& {van der Marel}(2005)}]{McLaughlin&vdMarel05}
{McLaughlin}, D.~E. \& {van der Marel}, R.~P. 2005, ApJS, 161, 304

\bibitem[{{Miller} \& {Lotz}(2007)}]{Miller&Lotz07}
{Miller}, B.~W. \& {Lotz}, J.~M. 2007, ApJ, 670, 1074

\bibitem[{{Miller} {et~al.}(1998){Miller}, {Lotz}, {Ferguson}, {Stiavelli}, \&
  {Whitmore}}]{Miller98}
{Miller}, B.~W., {Lotz}, J.~M., {Ferguson}, H.~C., {Stiavelli}, M., \&
  {Whitmore}, B.~C. 1998, ApJL, 508, L133

\bibitem[{{Muratov} \& {Gnedin}(2010)}]{Muratov&Gnedin10}
{Muratov}, A.~L. \& {Gnedin}, O.~Y. 2010, ArXiv:1002.1325

\bibitem[{{Olsen} {et~al.}(2004){Olsen}, {Miller}, {Suntzeff}, {Schommer}, \&
  {Bright}}]{Olsen04}
{Olsen}, K.~A.~G., {Miller}, B.~W., {Suntzeff}, N.~B., {Schommer}, R.~A., \&
  {Bright}, J. 2004, AJ, 127, 2674

\bibitem[{{O'Sullivan} {et~al.}(2007){O'Sullivan}, {Sanderson}, \&
  {Ponman}}]{O'Sullivan07}
{O'Sullivan}, E., {Sanderson}, A.~J.~R., \& {Ponman}, T.~J. 2007, MNRAS, 380,
  1409

\bibitem[{{Parmentier}(2009)}]{Parmentier09}
{Parmentier}, G. 2009, ArXiv:0901.3140

\bibitem[{{Parmentier} \& {Fritze}(2009)}]{Parmentier&Fritze09}
{Parmentier}, G. \& {Fritze}, U. 2009, ApJ, 690, 1112

\bibitem[{{Paturel} {et~al.}(2003){Paturel}, {Petit}, {Prugniel}, {Theureau},
  {Rousseau}, {Brouty}, {Dubois}, \& {Cambr{\'e}sy}}]{Paturel03}
{Paturel}, G., {Petit}, C., {Prugniel}, P., {et~al.} 2003, A\&A, 412, 45

\bibitem[{{Peng} {et~al.}(2006){Peng}, {Jord{\'a}n}, {C{\^o}t{\'e}},
  {Blakeslee}, {Ferrarese}, {Mei}, {West}, {Merritt}, {Milosavljevi{\'c}}, \&
  {Tonry}}]{Peng06}
{Peng}, E.~W., {Jord{\'a}n}, A., {C{\^o}t{\'e}}, P., {et~al.} 2006, ApJ, 639,
  95

\bibitem[{{Peng} {et~al.}(2008){Peng}, {Jordan}, {Cote}, {Takamiya}, {West},
  {Blakeslee}, {Chen}, {Ferrarese}, {Mei}, {Tonry}, \& {West}}]{Peng08}
{Peng}, E.~W., {Jordan}, A., {Cote}, P., {et~al.} 2008, astro-ph/0803.0330, 803

\bibitem[{{Pipino} {et~al.}(2007){Pipino}, {Puzia}, \& {Matteucci}}]{Pipino07}
{Pipino}, A., {Puzia}, T.~H., \& {Matteucci}, F. 2007, ApJ, 665, 295

\bibitem[{{Puzia} \& {Sharina}(2008)}]{Puzia&Sharina08}
{Puzia}, T.~H. \& {Sharina}, M.~E. 2008, ApJ, 674, 909

\bibitem[{{Renzini}(2006)}]{Renzini06}
{Renzini}, A. 2006, ARA\&A, 44, 141

\bibitem[{{Rhode} \& {Zepf}(2001)}]{Rhode&Zepf01}
{Rhode}, K.~L. \& {Zepf}, S.~E. 2001, AJ, 121, 210

\bibitem[{{Rhode} \& {Zepf}(2003)}]{Rhode&Zepf03}
{Rhode}, K.~L. \& {Zepf}, S.~E. 2003, AJ, 126, 2307

\bibitem[{{Rhode} \& {Zepf}(2004)}]{Rhode&Zepf04}
{Rhode}, K.~L. \& {Zepf}, S.~E. 2004, AJ, 127, 302

\bibitem[{{Rhode} {et~al.}(2007){Rhode}, {Zepf}, {Kundu}, \&
  {Larner}}]{Rhode07}
{Rhode}, K.~L., {Zepf}, S.~E., {Kundu}, A., \& {Larner}, A.~N. 2007, AJ, 134,
  1403

\bibitem[{{Rhode} {et~al.}(2005){Rhode}, {Zepf}, \& {Santos}}]{Rhode05}
{Rhode}, K.~L., {Zepf}, S.~E., \& {Santos}, M.~R. 2005, ApJL, 630, L21

\bibitem[{{Roberts} \& {Haynes}(1994)}]{Roberts&Haynes94}
{Roberts}, M.~S. \& {Haynes}, M.~P. 1994, ARA\&A, 32, 115

\bibitem[{{Ryan} {et~al.}(2007){Ryan}, {Hathi}, {Cohen}, {Malhotra}, {Rhoads},
  {Windhorst}, {Budav{\'a}ri}, {Pirzkal}, {Xu}, {Panagia}, {Moustakas}, {di
  Serego Alighieri}, \& {Yan}}]{Ryan07}
{Ryan}, R.~E.~J., {Hathi}, N.~P., {Cohen}, S.~H., {et~al.} 2007, ApJ, 668, 839

\bibitem[{{Schlegel} {et~al.}(1998){Schlegel}, {Finkbeiner}, \&
  {Davis}}]{Schlegel98}
{Schlegel}, D.~J., {Finkbeiner}, D.~P., \& {Davis}, M. 1998, ApJ, 500, 525

\bibitem[{{Seth} {et~al.}(2004){Seth}, {Olsen}, {Miller}, {Lotz}, \&
  {Telford}}]{Seth04}
{Seth}, A., {Olsen}, K., {Miller}, B., {Lotz}, J., \& {Telford}, R. 2004, AJ,
  127, 798

\bibitem[{{Sharina} {et~al.}(2010){Sharina}, {Chandar}, {Puzia}, {Goudfrooij},
  \& {Davoust}}]{Sharina10}
{Sharina}, M.~E., {Chandar}, R., {Puzia}, T.~H., {Goudfrooij}, P., \&
  {Davoust}, E. 2010, ArXiv:1002.2144

\bibitem[{{Sharina} {et~al.}(2005){Sharina}, {Puzia}, \& {Makarov}}]{Sharina05}
{Sharina}, M.~E., {Puzia}, T.~H., \& {Makarov}, D.~I. 2005, A\&A, 442, 85

\bibitem[{{Spitler} \& {Forbes}(2009)}]{Spitler09}
{Spitler}, L.~R. \& {Forbes}, D.~A. 2009, MNRAS, 392, L1

\bibitem[{{Spitler} {et~al.}(2008){Spitler}, {Forbes}, {Strader}, {Brodie}, \&
  {Gallagher}}]{Spitler08}
{Spitler}, L.~R., {Forbes}, D.~A., {Strader}, J., {Brodie}, J.~P., \&
  {Gallagher}, J.~S. 2008, MNRAS, 385, 361

\bibitem[{{Strader} {et~al.}(2006){Strader}, {Brodie}, {Spitler}, \&
  {Beasley}}]{Strader06}
{Strader}, J., {Brodie}, J.~P., {Spitler}, L., \& {Beasley}, M.~A. 2006, AJ,
  132, 2333

\bibitem[{{Strigari} {et~al.}(2008){Strigari}, {Bullock}, {Kaplinghat},
  {Simon}, {Geha}, {Willman}, \& {Walker}}]{Strigari08}
{Strigari}, L.~E., {Bullock}, J.~S., {Kaplinghat}, M., {et~al.} 2008, Nature,
  454, 1096

\bibitem[{{Tamura} {et~al.}(2006{\natexlab{a}}){Tamura}, {Sharples}, {Arimoto},
  {Onodera}, {Ohta}, \& {Yamada}}]{Tamura06a}
{Tamura}, N., {Sharples}, R.~M., {Arimoto}, N., {et~al.} 2006{\natexlab{a}},
  MNRAS, 373, 588

\bibitem[{{Tamura} {et~al.}(2006{\natexlab{b}}){Tamura}, {Sharples}, {Arimoto},
  {Onodera}, {Ohta}, \& {Yamada}}]{Tamura06b}
{Tamura}, N., {Sharples}, R.~M., {Arimoto}, N., {et~al.} 2006{\natexlab{b}},
  MNRAS, 373, 601

\bibitem[{{Tully} {et~al.}(2006){Tully}, {Rizzi}, {Dolphin}, {Karachentsev},
  {Karachentseva}, {Makarov}, {Makarova}, {Sakai}, \& {Shaya}}]{Tully06}
{Tully}, R.~B., {Rizzi}, L., {Dolphin}, A.~E., {et~al.} 2006, AJ, 132, 729

\bibitem[{{van den Bergh}(2000)}]{vdBergh00}
{van den Bergh}, S. 2000, PASP, 112, 932

\bibitem[{{van den Bergh} \& {Mackey}(2004)}]{vdBergh&Mackey04}
{van den Bergh}, S. \& {Mackey}, A.~D. 2004, MNRAS, 354, 713

\bibitem[{{Vesperini}(1998)}]{Vesperini98}
{Vesperini}, E. 1998, MNRAS, 299, 1019

\bibitem[{{Vesperini}(2000)}]{Vesperini00}
{Vesperini}, E. 2000, MNRAS, 318, 841

\bibitem[{{Walsh} {et~al.}(2007){Walsh}, {Jerjen}, \& {Willman}}]{Walsh07}
{Walsh}, S.~M., {Jerjen}, H., \& {Willman}, B. 2007, ApJL, 662, L83

\bibitem[{{Wehner} {et~al.}(2008){Wehner}, {Harris}, {Whitmore}, {Rothberg}, \&
  {Woodley}}]{Wehner08}
{Wehner}, E., {Harris}, B., {Whitmore}, B., {Rothberg}, B., \& {Woodley}, K.
  2008, astro-ph/0802.1723, 802

\bibitem[{{Weisz} {et~al.}(2008){Weisz}, {Skillman}, {Cannon}, {Dolphin},
  {Kennicutt}, {Lee}, \& {Walter}}]{Weisz08}
{Weisz}, D.~R., {Skillman}, E.~D., {Cannon}, J.~M., {et~al.} 2008, ApJ, 689,
  160

\bibitem[{{West} {et~al.}(1995){West}, {Cote}, {Jones}, {Forman}, \&
  {Marzke}}]{West95}
{West}, M.~J., {Cote}, P., {Jones}, C., {Forman}, W., \& {Marzke}, R.~O. 1995,
  ApJL, 453, L77+

\bibitem[{{White} \& {Rees}(1978)}]{White&Rees78}
{White}, S.~D.~M. \& {Rees}, M.~J. 1978, MNRAS, 183, 341

\bibitem[{{Whitmore} \& {Schweizer}(1995)}]{Whitmore&Schweizer95}
{Whitmore}, B.~C. \& {Schweizer}, F. 1995, AJ, 109, 960

\bibitem[{{Whitmore} {et~al.}(1999){Whitmore}, {Zhang}, {Leitherer}, {Fall},
  {Schweizer}, \& {Miller}}]{Whitmore99}
{Whitmore}, B.~C., {Zhang}, Q., {Leitherer}, C., {et~al.} 1999, AJ, 118, 1551

\bibitem[{{Willman} {et~al.}(2005{\natexlab{a}}){Willman}, {Blanton}, {West},
  {Dalcanton}, {Hogg}, {Schneider}, {Wherry}, {Yanny}, \&
  {Brinkmann}}]{Willman05}
{Willman}, B., {Blanton}, M.~R., {West}, A.~A., {et~al.} 2005{\natexlab{a}},
  AJ, 129, 2692

\bibitem[{{Willman} {et~al.}(2005{\natexlab{b}}){Willman}, {Dalcanton},
  {Martinez-Delgado}, {West}, {Blanton}, {Hogg}, {Barentine}, {Brewington},
  {Harvanek}, {Kleinman}, {Krzesinski}, {Long}, {Neilsen}, {Nitta}, \&
  {Snedden}}]{Willman05b}
{Willman}, B., {Dalcanton}, J.~J., {Martinez-Delgado}, D., {et~al.}
  2005{\natexlab{b}}, ApJL, 626, L85

\bibitem[{{Zepf} \& {Ashman}(1993)}]{Zepf&Ashman93}
{Zepf}, S.~E. \& {Ashman}, K.~M. 1993, MNRAS, 264, 611

\bibitem[{{Zucker} {et~al.}(2006{\natexlab{a}}){Zucker}, {Belokurov}, {Evans},
  {Kleyna}, {Irwin}, {Wilkinson}, {Fellhauer}, {Bramich}, {Gilmore}, {Newberg},
  {Yanny}, {Smith}, {Hewett}, {Bell}, {Rix}, {Gnedin}, {Vidrih}, {Wyse},
  {Willman}, {Grebel}, {Schneider}, {Beers}, {Kniazev}, {Barentine},
  {Brewington}, {Brinkmann}, {Harvanek}, {Kleinman}, {Krzesinski}, {Long},
  {Nitta}, \& {Snedden}}]{Zucker06}
{Zucker}, D.~B., {Belokurov}, V., {Evans}, N.~W., {et~al.} 2006{\natexlab{a}},
  ApJL, 650, L41

\bibitem[{{Zucker} {et~al.}(2006{\natexlab{b}}){Zucker}, {Belokurov}, {Evans},
  {Wilkinson}, {Irwin}, {Sivarani}, {Hodgkin}, {Bramich}, {Irwin}, {Gilmore},
  {Willman}, {Vidrih}, {Fellhauer}, {Hewett}, {Beers}, {Bell}, {Grebel},
  {Schneider}, {Newberg}, {Wyse}, {Rockosi}, {Yanny}, {Lupton}, {Smith},
  {Barentine}, {Brewington}, {Brinkmann}, {Harvanek}, {Kleinman}, {Krzesinski},
  {Long}, {Nitta}, \& {Snedden}}]{Zucker06b}
{Zucker}, D.~B., {Belokurov}, V., {Evans}, N.~W., {et~al.} 2006{\natexlab{b}},
  ApJL, 643, L103

\end{thebibliography}

\newpage\clearpage
\begin{deluxetable}{lccccccp{0.1cm}p{0.1cm}ccccccc}
\tablewidth{0pt}
\rotate
\tablecaption{Properties of the globular cluster systems of the studied dwarf galaxies\label{Table:SN}}
\tablehead{
\colhead{Galaxy} & 
\colhead{Type} & 
\colhead{$M_V$} & 
\colhead{$M/L_{V,{\rm Gal}}$} & 
\colhead{$M_{\star,{\rm V,Gal}}$} & 
\colhead{$M_{\rm HI}$} & 
\colhead{$N_{\rm blue}$} & 
\colhead{$N_{\rm tot}$} & 
\colhead{$M_{\rm blue}$} & 
\colhead{$M_{\rm tot}$} & 
\colhead{$S_{N, {\rm tot}}$} & 
\colhead{$S_{\rm L}$} &
\colhead{$S_{\rm M}$} &
\colhead{$\hat{T}$}  &
\colhead{$\log\eta_L$} & 
\colhead{$\log\eta_M$}\\ 
\colhead{} & 
\colhead{T-value} & 
\colhead{[mag]} & 
\colhead{} & 
\colhead{$10^{7}M_\odot$} & 
\colhead{$10^{7}M_\odot$} & 
\colhead{} & 
\colhead{} &
\colhead{$10^{4}M_\odot$} & 
\colhead{$10^{4}M_\odot$} & 
\colhead{} &
\colhead{} &
\colhead{} & 
\colhead{} & 
\colhead{} \\
\colhead{(1)} & 
\colhead{(2)} & 
\colhead{(3)} & 
\colhead{(4)} & 
\colhead{(5)} & 
\colhead{(6)} & 
\colhead{(7)} & 
\colhead{(8)} & 
\colhead{(9)} &
\colhead{(10)}&
\colhead{(11)}&
\colhead{(12)}&
\colhead{(13)}&
\colhead{(14)}&
\colhead{(15)}&
\colhead{(16)}
}
%\footnotesize
\startdata

%#c ID ch*10 %-10s 
%#c GTYPE d %3.1f 
%#c MGV d %7.2f 
%#c MLV d %8.3f 
%#c MstarG_1e6 d %9.2f 
%#c M_HI_LEDA_1E7 d %7.2f 
%#c NB i %5.0f 
%#c NTOT d %5.0f 
%#c MCLB_1E4 d %9.2f 
%#c MCLTOT_1E4 d %8.2f 
%#c SN_T d %7.2f 
%#c S_L d %8.2f 
%#c S_M d %8.2f 
%#c S_T d %8.2f 
%#c log_eta d %9.2f 
%#c log_eta_Sm d %9.2f 

D634-03 	&	$10.0	$\hspace{0.2cm} I	& $ -11.94$	&	0.691	&   0.35  &  0.49  &   0  &   1  &    0.00  &  10.82  & 16.75	&  1.14  &   1.29  & 119.05   & $ -4.39 $ & $ -4.77 $	 \\
DDO52   	&	$10.0	$\hspace{0.2cm} Im	& $ -14.98$	&	0.767	&   6.38  & 19.99  &   1  &   2  &   10.52  &  17.61  &  2.04	&  0.11  &   0.07  &   7.58   & $ -4.91 $ & $ -5.52 $	 \\
ESO\,059-01 	&	$10.0	$\hspace{0.2cm} IBsm	& $ -14.60$	&	0.867	&   5.08  &  8.26  &   1  &   1  &  143.93  & 143.93  &  1.45	&  1.31  &   1.08  &   7.50   & $ -3.90 $ & $ -4.32 $	 \\
ESO\,121-20 	&	$10.0	$\hspace{0.2cm} Im	& $ -13.64$	&	0.404	&   0.98  & 11.49  &   0  &   1  &    0.00  &	4.04  &  3.50	&  0.09  &   0.03  &   8.02   & $ -5.22 $ & $ -6.33 $	 \\
ESO\,137-18 	&	$5.0	$\hspace{0.2cm} SAsc	& $ -17.21$	&	1.049	&  68.04  & 34.14  &   4  &   7  &   69.69  & 111.20  &  0.91	&  0.09  &   0.11  &   6.85   & $ -4.65 $ & $ -4.82 $	 \\
ESO\,154-023	&	$8.9	$\hspace{0.2cm} SBsm	& $ -16.38$	&	0.662	&  19.99  & 81.70  &   0  &   3  &    0.00  &	3.52  &  0.84	&  0.01  &   0.00  &   2.95   & $ -5.94 $ & $ -6.65 $	 \\
ESO\,223-09 	&	$9.7	$\hspace{0.2cm} IAB	& $ -16.47$	&	 0.691  &   22.67 &  63.89 &	6 &    8 &   210.46 &  307.84 &   2.07  &   0.50 &    0.36 &	9.24  & $  -4.02$ & $  -4.61$	 \\
ESO\,269-58 	&	$-2.2	$\hspace{0.2cm}  I0	& $ -15.78$	&	 1.732  &   30.10 &   2.31 &	6 &    8 &    76.97 &	76.97 &   3.90  &   0.24 &    0.24 &   24.68  & $  -4.46$ & $  -4.49$	 \\
ESO\,269-66 	&	$-5.0	$\hspace{0.2cm}  dE,N	& $ -13.89$	&	 0.986  &    3.01 &   0.00 &	3 &    4 &   196.86 &  224.63 &  11.12  &   3.92 &    7.48 &  133.11  & $  -3.54$ & $  -3.54$	 \\
ESO\,274-01 	&	$6.6	$\hspace{0.2cm} Scd	& $ -17.47$	&	1.834	& 151.15  & 20.18  &   3  &  10  &   41.31  & 130.61  &  1.03	&  0.08  &   0.08  &   5.84   & $ -4.64 $ & $ -4.70 $	 \\
ESO\,349-031	&	$10.0	$\hspace{0.2cm}  IBm	& $ -11.87$	&	 0.629  &    0.30 &   1.34 &	0 &    1 &     0.00 &	 2.74 &  17.86  &   0.31 &    0.17 &   61.05  & $  -4.97$ & $  -5.71$	 \\
ESO\,381-20 	&	$9.8	$\hspace{0.2cm} IBsm	& $ -14.80$	&	 0.766  &    5.40 &  15.77 &	0 &    1 &     0.00 &	 2.28 &   1.20  &   0.02 &    0.01 &	4.72  & $  -5.75$ & $  -6.34$	 \\
ESO\,384-016	&	$-5	$\hspace{0.2cm} dSO/Im  & $ -13.72$	&	 0.691  &    1.80 &   0.50 &	0 &    2 &     0.00 &	 5.83 &   6.50  &   0.12 &    0.25 &   86.92  & $  -5.08$ & $  -5.19$	 \\
IC\,1959  	&	$8.4	$\hspace{0.2cm} SBsm	& $ -15.99$	&	0.672	&  14.17  & 18.73  &   4  &   7  &  159.26  & 172.34  &  2.81	&  0.44  &   0.52  &  21.28   & $ -4.16 $ & $ -4.53 $	 \\
IKN     	&	$?	$\hspace{0.2cm} dSph	& $ -11.51$	&	0.691	&   0.24  &  0.00  &   4  &   5  &   72.40  &  80.46  &124.44	& 12.57  &  34.24  &2127.66   & $ -3.41 $ & $ -3.41 $	 \\
KK 16    	&	$10.0	$\hspace{0.2cm}  I	& $ -12.38$	&	 0.691  &    0.52 &   0.69 &	0 &    1 &     0.00 &	 2.08 &  11.17  &   0.15 &    0.17 &   82.37  & $  -5.21$ & $  -5.58$	 \\
KK\,17    	&	$10.0	$\hspace{0.2cm}  I	& $ -10.57$	&	 0.691  &    0.10 &   0.55 &	0 &    1 &     0.00 &	 1.09 &  59.16  &   0.41 &    0.17 &  154.08  & $  -5.06$ & $  -5.87$	 \\
KK\,197   	&	$10.0	$\hspace{0.2cm}  Im	& $ -13.04$	&	 0.691  &    0.96 &   0.00 &	2 &    3 &   148.96 &  151.97 &  18.24  &   5.80 &   15.78 &  311.53  & $  -3.51$ & $  -3.51$	 \\
KK\,246   	&	$10.0	$\hspace{0.2cm}  I	& $ -13.77$	&	 0.354  &    0.97 &  11.92 &	2 &    2 &    41.04 &	41.04 &   6.21  &   0.80 &    0.32 &   15.52  & $  -4.25$ & $  -5.37$	 \\
KK\,27    	&	$10.0	$\hspace{0.2cm}  I	& $ -10.14$	&	 0.691  &    0.07 &   0.00 &	0 &    2 &     0.00 &	 6.03 & 175.80  &   3.33 &    9.00 & 2985.07  & $  -4.21$ & $  -4.21$	 \\
KKH\,77   	&	$10.0	$\hspace{0.2cm}  I	& $ -14.58$	&	 0.691  &    3.98 &   4.67 &	1 &    2 &    24.78 &	27.11 &   2.94  &   0.25 &    0.31 &   23.13  & $  -4.62$ & $  -4.96$	 \\
KKS\,55   	&	$-	$\hspace{0.2cm} dSph	& $ -11.17$	&	0.691	&   0.17  &  0.00  &   1  &   1  &   14.00  &  14.00  & 34.04	&  2.99  &   8.14  & 581.40   & $ -4.09 $ & $ -4.09 $	 \\
NGC\,1311   	&	$8.7	$\hspace{0.2cm} SBm	& $ -15.76$	&	0.691	&  11.79  &  8.74  &   3  &   5  &   63.64  &  68.31  &  2.48	&  0.21  &   0.33  &  24.36   & $ -4.51 $ & $ -4.75 $	 \\
NGC\,247    	&	$6.9	$\hspace{0.2cm} SABd	& $ -18.76$	&	0.860	& 232.54  &137.06  &   1  &  25  &   14.80  & 398.20  &  0.78	&  0.08  &   0.11  &   6.76   & $ -4.50 $ & $ -4.70 $	 \\
NGC\,4163   	&	$9.9	$\hspace{0.2cm} I	& $ -14.21$	&	 0.469  &    1.92 &   1.42 &	2 &    2 &   111.57 &  111.57 &   4.14  &   1.45 &    3.34 &   59.90  & $  -3.92$ & $  -4.16$	 \\
NGC\,4605   	&	$5.0	$\hspace{0.2cm} SBd	& $ -18.41$	&	0.883	& 172.97  & 25.88  &   7  &  22  &   85.92  & 350.41  &  0.95	&  0.10  &   0.18  &  11.06   & $ -4.46 $ & $ -4.52 $	 \\
NGC\,5237   	&	$-	$\hspace{0.2cm} I?	& $ -15.45$	&	 1.211  &   15.53 &   3.10 &	2 &    3 &    47.32 &	53.21 &   1.98  &   0.22 &    0.29 &   16.10  & $  -4.54$ & $  -4.62$	 \\
NGC\,784    	&	$7.8	$\hspace{0.2cm} SBdm	& $ -16.87$	&	1.785	&  84.65  & 28.90  &   4  &   6  &   36.70  &  40.71  &  1.07	&  0.05  &   0.04  &   5.28   & $ -5.00 $ & $ -5.13 $	 \\
UGC\,1281   	&	$7.5	$\hspace{0.2cm} Sdm	& $ -15.30$	&	1.134	&  12.67  & 16.49  &   1  &   2  &   17.30  &  31.95  &  1.52	&  0.15  &   0.11  &   6.86   & $ -4.73 $ & $ -5.09 $	 \\
UGC\,3755   	&	$9.9	$\hspace{0.2cm} Im	& $ -15.50$	&	 0.691  &    9.28 &  13.95 &	6 &    9 &   174.20 &  239.30 &   5.68  &   0.95 &    1.03 &   38.74  & $  -3.90$ & $  -4.30$	 \\
UGC\,3974   	&	$9.8	$\hspace{0.2cm} IBm	& $ -15.33$	&	 0.691  &    7.93 &  63.75 &	4 &    4 &   118.66 &  118.66 &   2.95  &   0.55 &    0.17 &	5.58  & $  -4.01$ & $  -4.97$	 \\
UGC\,4115   	&	$9.9	$\hspace{0.2cm} Im	& $ -15.12$	&	 0.900  &    8.52 &  20.95 &	1 &    5 &    17.79 &	29.86 &   4.48  &   0.17 &    0.10 &   16.97  & $  -4.71$ & $  -5.25$	 \\
UGC\,685    	&	$9.2	$\hspace{0.2cm} Im	& $ -14.35$	&	 0.691  &    3.22 &   5.88 &	5 &    5 &   136.93 &  136.93 &   9.10  &   1.56 &    1.51 &   54.96  & $  -3.86$ & $  -4.32$	 \\
UGC\,7369   	&	$-3	$\hspace{0.2cm} dSph	& $ -16.17$	&	 0.691  &   17.20 &   0.00 &   14 &   22 &  1208.08 & 1255.32 &   7.49  &   2.68 &    7.30 &  127.92  & $  -3.34$ & $  -3.34$	 \\
UGC\,8638   	&	$9.9	$\hspace{0.2cm} Im	& $ -13.69$	&	 0.641  &    1.63 &   1.17 &	2 &    3 &    26.27 &  246.14 &  10.03  &   5.16 &    8.81 &  107.33  & $  -3.45$ & $  -3.69$	 \\
UGC\,8760   	&	$9.8	$\hspace{0.2cm} IBm	& $ -13.16$	&	 0.691  &    1.08 &   1.86 &	0 &    1 &     0.00 &	 1.33 &   5.45  &   0.05 &    0.05 &   34.07  & $  -5.59$ & $  -6.03$	 \\
UGCA 86    	&	$9.9	$\hspace{0.2cm} Im	& $ -16.13$	&	 0.691  &   16.58 &  48.25 &   10 &   11 &   685.29 & 1148.91 &   3.89  &   2.55 &    1.77 &   16.97  & $  -3.37$ & $  -3.96$	 \\
UGCA 92    	&	$10.0	$\hspace{0.2cm} Im	& $ -14.71$	&	 0.691  &    4.48 &   7.62 &	1 &    2 &    19.87 &	46.06 &   2.61  &   0.38 &    0.38 &   16.53  & $  -4.42$ & $  -4.86$	 \\
LMC     	&	$9.1	$\hspace{0.2cm} I	& $ -18.36$	&	 0.870  &  162.75 &  44.12 &   16 &   16 &   330.27 &  330.27 &   0.72  &   0.09 &    0.16 &	7.73  & $  -4.47$ & $  -4.57$	 \\
SMC     	&	$8.9	$\hspace{0.2cm} I	& $ -16.82$	&	 0.815  &   36.91 &  51.20 &	1 &    1 &    32.16 &	32.16 &   0.19  &   0.04 &    0.04 &	1.13  & $  -5.09$ & $  -5.47$	 \\
NGC\,1427A  	&	$9.9	$\hspace{0.2cm} IBsm	& $ -18.50$	&	 0.937  &  199.41 & 176.63 &   38 &   38 &   605.26 &  605.26 &   1.66  &   0.15 &    0.16 &   10.11  & $  -4.24$ & $  -4.52$	 \\

\enddata	

\tablecomments{Columns (1) through (6) contain the galaxy ID, morphological classification as listed in HyperLEDA and NED, its absolute magnitude, $V-$band mass-to-light ratio (see Sect.\,\ref{SmSect}), stellar mass and {\sc HI} mass (see Sect.\,\ref{SmSect}), respectively. In columns (7) through (10) is the number of "blue" GCs $0.7\!<\!V\!-\!I\!<\!1.0$\,mag, the total number of GCs in the galaxy, the mass of the blue GCs and the total mass of the GC system, respectively. In column (11) is the GC specific frequency, in (12) specific luminosity, (13) specific mass and in (14) number of GCs normalized to the galaxy mass. The logarithm of the luminosity and mass normalized GC formation efficiencies are shown in columns (15) and (16), respectively (see Sect.\,\ref{Sect:GCefficiency}).}

\end{deluxetable}

\newpage\clearpage
\appendix
\section{Scaling relations of GC systems as a function of galaxy mass 
(dark and baryon) and luminosity}\label{appendix}

\subsection{GCSs scaling relations in the "SNe feedback" galaxy mass regime}
Using Equations\,(\ref{eqn:Mgcs=etaMh}) through (\ref{eqn:kappa}) we express 
Equations\,(\ref{eqn:sm}), (\ref{eqn:sl}), and (\ref{eqn:T}) as a function 
of galaxy luminosity, total halo and baryon mass for ${\cal M}_h\!<\!3\times10^{10}{\cal M}_\odot$.

\begin{equation}
\log S_N = 6+0.4M_{V,\odot} + \log\left(\frac{\eta\kappa_1^{0.6}}{m_{\rm TO}}\right) - 0.4\log L_V
\label{eqn:SNevolLv}
\end{equation}
\begin{equation}
\hspace*{1cm} = 6+0.4M_{V,\odot} + \log\left(\frac{\eta\kappa_1}{m_{\rm TO}}\right) - \frac{2}{3}\log {\cal M}_h
\label{eqn:SNevolMh}
\end{equation}
\begin{equation}
\hspace*{1cm} = 6+0.4M_{V,\odot} + \log\left(\frac{\eta}{m_{\rm TO}}\frac{\kappa_1^{0.6}}{\Upsilon_V^{0.6}}\right) - 0.4\log {\cal M}_{\rm b}
\label{eqn:SNevolMb}
\end{equation}

\begin{equation}
\log S_L = 2 + \log\left(\frac{\eta\kappa_1^{0.6}}{\gamma_V}\right) - 0.4\log L_V
\label{eqn:slevolLv}
\end{equation}
\begin{equation}
\hspace*{0.99cm} = 2 + \log\left(\frac{\eta\kappa_1}{\gamma_V}\right) - \frac{2}{3}\log {\cal M}_h
\label{eqn:slevolMh}
\end{equation}
\begin{equation}
\hspace*{0.99cm} = 2 + \log\left(\frac{\eta\kappa_1^{0.6}\Upsilon^{0.4}_V}{\gamma_V}\right) - 0.4\log {\cal M}_{\rm b}
\label{eqn:slevolMb}
\end{equation}

\begin{equation}
\log S_M = 2 + \log\left(\frac{\eta\kappa_1^{0.6}}{\Upsilon_V}\right) - 0.4\log L_V
\label{eqn:smevolLv}
\end{equation}
\begin{equation}
\hspace*{1.05cm} = 2 + \log\left(\frac{\eta\kappa_1}{\Upsilon_V}\right) - \frac{2}{3}\log {\cal M}_h
\label{eqn:smevolMh}
\end{equation}
\begin{equation}
\hspace*{1.05cm} = 2 + \log\left(\eta\frac{\kappa_1^{0.6}}{\Upsilon_V^{0.6}}\right) - 0.4\log {\cal M}_{\rm b}
\label{eqn:smevolMb}
\end{equation}

\begin{equation}
\log \hat{T} = 9 + \log\left(\frac{\eta\kappa_1^{0.6}}{\Upsilon_V m_{\rm TO}}\right) - 0.4\log L_V
\label{eqn:tevolLv}
\end{equation}
\begin{equation}
\hspace*{0.8cm} = 9 + \log\left(\frac{\eta\kappa_1}{\Upsilon_V m_{\rm TO}}\right) - \frac{2}{3}\log {\cal M}_h,
\label{eqn:tevolMh}
\end{equation}
\begin{equation}
\hspace*{0.8cm} = 9 + \log\left(\frac{\eta\kappa_1^{0.6}}{m_{\rm TO}\Upsilon_V^{0.6}}\right) - 0.4\log{\cal M}_{\rm b},
\label{eqn:tevolMb}
\end{equation}

\subsection{GCSs scaling relations in the "Virial-shock" galaxy mass regime}
Using Equations\,(\ref{eqn:Mgcs=etaMh}) through (\ref{eqn:kappa2}) we express 
Equations\,(\ref{eqn:sm}), (\ref{eqn:sl}), and (\ref{eqn:T}) as a function 
of galaxy luminosity, total halo and baryon mass for ${\cal M}_h>3\times10^{10}{\cal M}_\odot$.

\begin{equation}
\log S_N  = 6+0.4M_{V,\odot} + \log\left(\frac{\eta\kappa_2^{2}}{m_{\rm TO}}\right) +1.2\log L_V
\label{eqn:SNevolLv2}
\end{equation}
\begin{equation}
\hspace*{1cm} = 6+0.4M_{V,\odot} + \log\left(\frac{\eta\kappa_2^{-0.8}}{m_{\rm TO}}\right) + \frac{3}{5}\log {\cal M}_h
\label{eqn:SNevolMh2}
\end{equation}
\begin{equation}
\hspace*{1cm} = 6+0.4M_{V,\odot} + \log\left(\frac{\eta\kappa_2^{2}}{m_{\rm TO}\Upsilon^{1.2}_V}\right) +1.2\log {\cal M}_{\rm b}
\label{eqn:SNevolMb2}
\end{equation}

\begin{equation}
\log S_L = 2 + \log\left(\frac{\eta\kappa_2^{2}}{\gamma_V}\right) +\log L_V
\label{eqn:slevolLv2}
\end{equation}
\begin{equation}
\hspace*{0.98cm} = 2 + \log\left(\frac{\eta\kappa_2}{\gamma_V}\right) + \frac{1}{2}\log {\cal M}_h
\label{eqn:slevolMh2}
\end{equation}
\begin{equation}
\hspace*{0.98cm} = 2 + \log\left(\frac{\eta\kappa_2^{2}}{\gamma_V\Upsilon_V}\right) +\log {\cal M}_{\rm b}
\label{eqn:slevolMb2}
\end{equation}

\begin{equation}
\log S_M = 2 + \log\left(\frac{\eta\kappa_2^{2}}{\Upsilon_V}\right) +\log L_V
\label{eqn:smevolLv2}
\end{equation}
\begin{equation}
\hspace*{1.05cm} = 2 + \log\left(\frac{\eta\kappa_2}{\Upsilon_V}\right) + \frac{1}{2}\log {\cal M}_h
\label{eqn:smevolMh2}
\end{equation}
\begin{equation}
\hspace*{1.05cm} = 2 + \log\left(\frac{\eta\kappa_2^{2}}{\Upsilon_V^{2}}\right) +\log {\cal M}_{\rm b}
\label{eqn:smevolMb2}
\end{equation}

\begin{equation}
\log \hat{T} = 9 + \log\left(\frac{\eta\kappa_2^{2}}{\Upsilon_V m_{\rm TO}}\right) +\log L_V
\label{eqn:tevolLv2}
\end{equation}
\begin{equation}
\hspace*{0.8cm} = 9 + \log\left(\frac{\eta\kappa_2}{\Upsilon_V m_{\rm TO}}\right) + \frac{1}{2}\log {\cal M}_h,
\label{eqn:tevolMh2}
\end{equation}
\begin{equation}
\hspace*{0.8cm} = 9 + \log\left(\frac{\eta}{m_{\rm TO}}\frac{\kappa_2^{2}}{\Upsilon_V^{2}}\right) +\log{\cal M}_{\rm b},
\label{eqn:tevolMb2}
\end{equation}

\subsection{Continuous GCSs scaling relations from the "Virial-shock" to the "SNe feedback" galaxy mass regime}
\begin{equation}
S_N  = 10^{6+0.4M_{V,\odot}}\frac{\eta}{m_{\rm TO}}\left(\kappa_1^{0.6}L_V^{-0.4} +\kappa_2^{2}L_V^{1.2}\right)
\label{eqn:CSNevolLv2}
\end{equation}
\begin{equation}
\hspace*{0.55cm} = 10^{6+0.4M_{V,\odot}}\frac{\eta}{m_{\rm TO}}\left(\kappa_1{\cal M}_h^{-2/3}+\kappa_2^{-0.8}{\cal M}_h^{3/5}\right)
\label{eqn:CSNevolMh2}
\end{equation}
\begin{equation}
\hspace*{0.55cm} = 10^{6+0.4M_{V,\odot}}\frac{\eta}{m_{\rm TO}}\left((\kappa_1\Upsilon_V)^{0.6}{\cal M}_{\rm b}^{-0.4}+\frac{\kappa_2^{2}}{\Upsilon_V^{1.2}}{\cal M}_{\rm b}^{1.2}\right)
\label{eqn:CSNevolMb2}
\end{equation}

\begin{equation}
S_L = 10^2\frac{\eta}{\gamma_V}\left(\kappa_1^{0.6}L_V^{-0.4} +\kappa_2^{2}L_V\right)
\label{eqn:CSLevolLv2}
\end{equation}
\begin{equation}
\hspace*{0.53cm} = 10^2\frac{\eta}{\gamma_V}\left(\kappa_1{\cal M}_h^{-2/3}+\kappa_2{\cal M}_h^{1/2}\right)
\label{eqn:CSLevolMh2}
\end{equation}
\begin{equation}
\hspace*{0.53cm} = 10^2\frac{\eta}{\gamma_V}\left(\kappa_1^{0.6}\Upsilon_V^{0.4}{\cal M}_{\rm b}^{-0.4}+\frac{\kappa_2^2}{\Upsilon_V}{\cal M}_{\rm b}\right)
\label{eqn:CSLevolMb2}
\end{equation}

\begin{equation}
S_M = 10^2\frac{\eta}{\Upsilon_V}\left(\kappa_1^{0.6}L_V^{-0.4}+\kappa_2^2L_V\right)
\label{eqn:CSMevolLv2}
\end{equation}
\begin{equation}
\hspace*{0.6cm} = 10^2\frac{\eta}{\Upsilon_V}\left(\kappa_1{\cal M}_h^{-2/3}+\kappa_2{\cal M}_h^{1/2}\right)
\label{eqn:CSMevolMh2}
\end{equation}
\begin{equation}
\hspace*{0.6cm} = 10^2\eta\left(\left(\frac{\kappa_1}{\Upsilon_V}\right)^{0.6}{\cal M}_{\rm b}^{-0.4}+\left(\frac{\kappa_2}{\Upsilon_V}\right)^2{\cal M}_{\rm b}\right)
\label{eqn:CSMevolMb2}
\end{equation}

\begin{equation}
\hat{T} = 10^9\frac{\eta}{m_{\rm TO}\Upsilon_V}\left(\kappa_1^{0.6}L_V^{-0.4}+\kappa_2^2L_V\right)
\label{eqn:CTevolLv2}
\end{equation}
\begin{equation}
\hspace*{0.33cm} = 10^9\frac{\eta}{m_{\rm TO}\Upsilon_V}\left(\kappa_1{\cal M}_h^{-2/3}+\kappa_2{\cal M}_h^{1/2}\right)
\label{eqn:CTevolMh2}
\end{equation}
\begin{equation}
\hspace*{0.33cm} = 10^9\frac{\eta}{m_{\rm TO}}\left(\left(\frac{\kappa_2}{\Upsilon_V}\right)^{0.4}{\cal M}_{\rm b}^{-0.4}+\left(\frac{\kappa_2}{\Upsilon_V}\right)^2{\cal M}_{\rm b}\right)
\label{eqn:CTevolMb2}
\end{equation}

\end{document}